\documentclass[a4paper]{article}
\usepackage[utf8]{inputenc}
\usepackage[english]{babel}
\usepackage{amsfonts}
\usepackage{amssymb}
\usepackage{amsthm}
\usepackage[fleqn]{nccmath}
\usepackage{amsmath}
\usepackage[sc]{mathpazo}
\usepackage{mathtools}
\usepackage[titletoc, header, page]{appendix}
\usepackage{pdfpages}   
\usepackage{array}      
\usepackage{booktabs}   
\usepackage{longtable}  
\usepackage[pdfusetitle]{hyperref}
\usepackage{graphicx} 
\usepackage{epstopdf}
\usepackage{multirow}
\usepackage[Conny]{fncychap}
\usepackage{makecell}
\usepackage{spreadtab}
\usepackage{soul}
\usepackage{xfrac}
\usepackage{flexisym}
\usepackage{keyval}
\usepackage{tabularx,blindtext}
\usepackage{siunitx}
\usepackage{subfig}
\usepackage{textcomp}
\usepackage{caption}
\usepackage{calc}
\usepackage{rotating}
\usepackage{mwe}
\usepackage{float}
\usepackage{empheq}
\usepackage{cite}
\usepackage{physics}
\usepackage{pict2e}
\usepackage{amsopn}
\usepackage{subfig}
\usepackage{diagbox}
\usepackage{fancyhdr}
\usepackage{empheq}
\usepackage[flushleft]{threeparttable}
\usepackage{setspace}   

\usepackage{geometry}            
\numberwithin{equation}{section}

\usepackage{lineno}

\begin{document}

\title{Less Is More in Chemotherapy of Breast Cancer
\thanks{Emails: f.ansarizadeh@deakin.edu.au (F. Ansarizadeh), tonghuazhang@swin.edu.au (T. Zhang)}
\author{Fatemeh Ansarizadeh\thanks{Corresponding author}, Tonghua Zhang.}
\\
{\small School of IT, Faculty of Eng., Sci. and the Built Env., Deakin University, Waurn Ponds, Victoria 3216, Australia 
 Department of Mathematics, Swinburne University of Technology, Hawthorn, Victoria 3122, Australia}}
\date{}
\maketitle 

{\noindent\bf Abstract:} 
This study presents a mathematical model that captures the interactions among tumor cells, healthy cells, and immune cells in a tumor-bearing host, with a specific focus on breast cancer. Incorporating the concept of delay, the model consists of four differential equations to analyze these cellular dynamics. The findings demonstrate the superior efficacy of metronomic chemotherapy compared to the maximum tolerated dose (MTD) method and underscore the necessity of adjunct therapies. Oscillatory tumor cell dynamics revealed by the model highlight the challenges of achieving complete tumor elimination through chemotherapy alone. Sensitivity analysis confirms the robustness of the model, particularly under metronomic treatment protocols, aligning with experimental observations regarding metronomic-to-MTD dosage ratios. Furthermore, the results emphasize the importance of synergistic effects from combination therapies. This biologically consistent framework provides valuable insights into tumor-immune interactions and offers a foundation for optimizing therapeutic strategies in cancer treatment.

\vskip0.2cm \noindent{\bf Keywords:} Breast cancer, Cell cycle, Chemotherapy, Delay Mathematical modelling

\section{Introduction}\label{int}
Breast cancer remains the most commonly diagnosed cancer and the leading cause of cancer-related mortality among females \cite{sha2024global}. This trend is not recent, as historical data indicate a consistent pattern over the previous decades \cite{mcpherson2000abc}. While predominantly affecting females, breast cancer can also occur in males, with its incidence among men showing a significant rise from 1990 to 2017 \cite{bhardwaj2024male}.
Interestingly, the highest incidence rates of breast cancer are observed in North America and Northern Europe, whereas less developed regions in Asia and Africa report the lowest rates \cite{richie2003breast}. Even more unexpectedly, studies have shown that the incidence of breast cancer increases in successive generations of Asian immigrants to high-risk areas. This suggests that lifestyle and environmental factors may have a greater influence on the development of breast cancer than genetic predispositions \cite{costanza2008epidemiology}. 
A recent study reveals a significant increase in the burden of breast cancer by 2050 \cite{liao2025inequality},
which serves as a stark warning about the urgent need to develop more effective treatment strategies for this life-threatening disease \cite{siegel2017cancer}.  

Chemotherapy remains the most commonly employed treatment for breast cancer and is administered through various protocols \cite{aprile2021hypertermic}. Among these, the Maximum Tolerated Dose (MTD) and metronomic protocols hold particular significance in clinical practice \cite{cazzaniga2021metronomic}.
MTD protocols involve administering the highest dose of chemotherapy that patients can tolerate without experiencing severe toxicity, with the goal of maximizing the eradication of tumor cells. In contrast, metronomic chemotherapy involves the continuous administration of lower doses with minimal breaks, focusing on targeting tumor angiogenesis while minimizing side effects \cite{bandini2024metronomic, basar2024optimizing}. Understanding the dynamics of these protocols and their impact on tumor progression is crucial for optimizing therapeutic strategies and enhancing patient outcomes.

Mathematical modeling provides a powerful tool to investigate these dynamics, offering valuable insights into optimizing treatment strategies and mitigating the global impact of breast cancer. 
Significant research has been dedicated to modeling the interactions between tumor and immune cells to better understand tumor responsiveness to treatment and to establish reasonable expectations for the survival of tumor-bearing hosts \cite{ ansarizadeh2017modelling}. Despite extensive efforts and substantial investments in studying tumor-immune dynamics, the precise roles and mechanisms of immune cells in tumor regression remain inadequately understood \cite{xu2021technological}.
Many existing models often oversimplify these interactions, for example, by neglecting competition terms among cell types or disregarding intrinsic delays in the tumor cell life cycle. Such simplifications may limit the capacity of these models to fully capture the complexities inherent in tumor-immune interactions. 

To address these limitations, our study extends previous modeling efforts by incorporating all competition terms among various cell types and explicitly accounting for delays in the tumor cell life cycle. Unlike many earlier approaches that primarily rely on ordinary differential equations (ODEs) \cite{enderling2007mathematical,  roe2011mathematical}, this work aims to provide a more comprehensive framework for analyzing the complex dynamics of tumor-immune interactions.
The parameters of the proposed system were derived primarily from clinical and experimental literature to ensure biological relevance and consistency. In cases where no precedents were available, rational conjectures were employed to assign reasonable parameter values, guided by biological interpretations and theoretical justifications. This approach ensures that the model maintains a balance between empirical grounding and flexibility, thus allowing it to capture the complexity of tumor-immune interactions while remaining applicable to real-world scenarios.
The proposed model is analyzed under two scenarios: a healthy host and a compromised host. The theoretical results demonstrate that in a healthy host, the immune system can successfully eradicate tumor cell mutations. In contrast, under conditions of immune system compromise, tumor cells evade immune surveillance, continue to proliferate, and ultimately form a tumor.
Additionally, the model highlights the superior efficacy of the metronomic chemotherapy approach over the Maximum Tolerated Dose (MTD) protocol. It emphasizes the necessity of adjunct therapies by revealing oscillatory dynamics in the tumor cell population, underscoring that chemotherapy alone is insufficient for complete tumor eradication.
Sensitivity analysis further confirms the robustness of the model, particularly under metronomic treatment protocols. These results validate the model’s potential to optimize therapeutic strategies and offer valuable insights for enhancing the efficacy of cancer treatment regimens.

The remainder of this paper is organized as follows: In Section \ref{Biological background}, a brief biological background is provided to establish the foundation necessary for understanding this research. The proposed mathematical model is introduced in Section \ref{Fundamentals of the mathematical model}, where its structure and underlying assumptions are detailed. Section \ref{ex1} presents extensive {\it{in silico}} experiments conducted to evaluate the performance of the model. The parameters of the model are clarified in Section \ref{Parameter set for the proposed model}, along with their biological interpretations. In Section \ref{Model analysis}, equilibrium points are analyzed for both tumor-free and tumor-bearing hosts, followed by a stability analysis that incorporates the effect of delay. The critical role of the immune system in tumor regression and progression is addressed in Section \ref{Effectiveness of the immune system}. Section \ref{Treatments strategies} compares different therapeutic strategies, including metronomic chemotherapy and the Maximum Tolerated Dose (MTD) protocol. Section \ref{Sensitivity analysis of the proposed model} presents a sensitivity analysis of the model, accompanied by mathematical analyses of stability, immune system influence, and parameter sensitivity. Finally, the results are compared with previous studies to verify the authenticity and reliability of the findings.

\section{Delay in cell cycle }\label{Biological background}

Time-delay systems are prevalent not only in scientific research but also in engineering problems, as they yield more realistic results \cite{wu2015time, niculescu2012advances}. Delay can be either a constant, known value—introducing less complexity to the system—or a distributed delay, which may adversely affect stability. In this study, we assume that the time delay is a constant value. Specifically, in the context of cell interactions, we propose that the time lag is attributed to the maturation of tumor cells. This implies that the primary source of delay is the time required for tumor cells to progress into the mitosis phase. A more detailed explanation of this concept is provided in Section \ref{Depicting the cell cycle}.

\subsection{Depicting the cell cycle} \label{Depicting the cell cycle}

\begin{figure}[ht]
\centering
\includegraphics[scale=0.3]{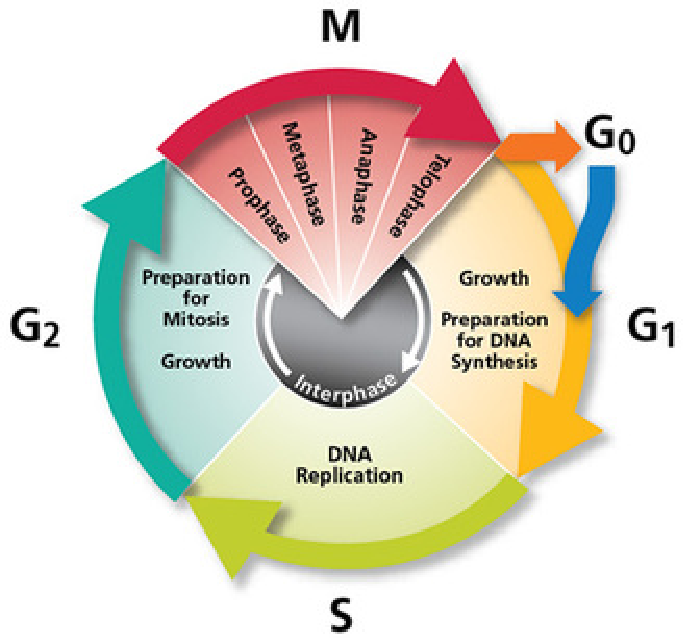}
\caption { Mammalian cell cycle: Interphase comprises $G_0$ to $G_2$, while prophase, metaphase, anaphase, and telophase form mitosis.
Mitosis is assumed to occur instantly and interphase lasts considerably longer.}
\label{cellcycle}
\end{figure}

A cell cycle encompasses all processes leading to cell growth and division, divided into five phases: interphase, prophase, metaphase, anaphase, and telophase \cite{alonso2024decoding}. Among these phases, interphase takes significantly longer compared to the others. Interphase consists of four stages, progressing from the $G_0$ phase to the $G_2$ phase, as illustrated in Figure \ref{cellcycle}.
In the $G_0$ phase, also known as the quiescent phase, cells are alive but not actively growing or preparing for division, essentially remaining dormant. The next phase, $G_1$, marks the initiation of cell growth as cells begin preparing for DNA synthesis. During the $S$ phase, DNA replication occurs, followed by the $G_2$ phase, where cells continue to grow and mature in preparation for mitosis. The subsequent phases—prophase, metaphase, anaphase, and telophase—constitute the $M$ (mitosis) phase, which culminates in cell division.

In Section \ref{Depicting cell cycle by equations}, a simplified yet practical mathematical model is employed to illustrate the kinetics of the cell cycle, with a specific emphasis on the proliferation and division of tumor cells.

\subsection{Mathematical interpretation of the cell cycle}\label{Depicting cell cycle by equations}

We modify our existing model \cite{ansarizadeh2017modelling} to enhance its predictability by incorporating the intrinsic delay in the cell cycle. The equations governing the behavior of normal and immune cells remain the same as in the original model, without considering the delay effect. However, since mitosis in tumor cells is longer and more complex, the dynamics of this phase for tumor cells are described in detail. 

\begin{figure}[ht]
\centering
\includegraphics[scale=0.6]{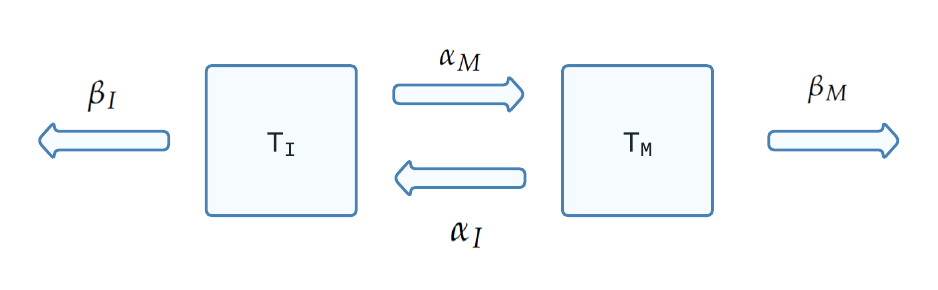}
\caption{Cell cycle dynamics of tumor cells: Tumor cells split into two daughter cells, starting with the interphase period of the next cycle at a rate of $\alpha_I$. The transition rate of cells between interphase and mitosis is represented by $\alpha_{\it{i}}$. The death rate of cells in each phase is denoted by $\beta_{\it{i}}$.}
\label{schematic}
\end{figure}

Initially, we model the interactions among different types of cells in the absence of chemotherapeutic agents to verify and validate the model. Once the model has been benchmarked, we simulate the beneficial effects of chemotherapy.
To achieve this, we adopt the well-established prey-predator model and extend it to a prey-predator-protector model, where normal cells, tumor cells, and immune cells (along with interleukin, secreted by white blood cells) are represented as prey, predator, and protector species, respectively.
To explain tumor cell dynamics, we must model all the prerequisite stages of the cell cycle, including $G_0$ (which is negligibly short in tumor cells), $G_1$, $S$, and $G_2$, which together form the interphase period. Tumor cells in any of these interphase stages are represented by $T_I$, while tumor cells in the mitosis stage are represented by $T_M$.
The dynamics between these two categories of tumor cells are illustrated in Figure \ref{schematic}, where $\alpha$ and $\beta$ represent the growth and death rates, respectively.

Equations \eqref{1} and \eqref{2} are the mathematical representation of this dynamics shown in Figure \ref{schematic}:
\begin{align}
\label{1}
\dv{ T_I(t)}{ t}&=2\alpha_I T_M(t)-\alpha_MT_I(t)-\beta_IT_I(t),\\
\label{2}
\dv{T_M(t)}{t}&=\alpha_M T_I(t)-\alpha_IT_M(t)-\beta_MT_M(t).
\end{align}

The coefficient 2 in equation \eqref{1} justifies the splitting of a parent cell into two daughter cells, initiating a new cycle. As shown in Figure \ref{cellcycle}, it takes a longer interval for tumor cells to experience the interphase stage, while the four phases that form mitosis occur instantly. Hence, if the duration of interphase is denoted by $\tau$, representing the delay, $T_I(t)$ is replaced by $T_I(t-\tau)$ in the equation that governs this process. Consequently, equations \eqref{3} and \eqref{4} can incorporate the inherent delay in the cell cycle into the dynamics of tumor cells.
\begin{align} 
\label{3}
\dv{ T_I(t)}{ t}&=2 \alpha_I T_M(t)-\alpha_MT_I(t-\tau)-\beta_IT_I(t),\\
\label{4}
\dv{T_M(t)}{t}&=\alpha_M T_I(t-\tau)-\alpha_IT_M(t)-\beta_MT_M(t).
\end{align}
It is important to note that the delay applies specifically to the representation of cell cycle dynamics within tumor cells across various phases. However, when describing the interactions between tumor cells and other cell types, it is essential to consider the populations of cells at the same moment. Given that the growth of various cell types shares similar qualitative characteristics, the system outlined in Section \ref{Fundamentals of the mathematical model} was chosen as a representative model for the tumor cell cycle.

\section{Fundamentals of the mathematical model}\label{Fundamentals of the mathematical model}

Here, we extend the Lotka-Volterra prey-predator model \cite{merlo2006cancer}, which describes species competition for survival, to incorporate predator-prey-protector dynamics within tumor-immune interactions. 

\subsection{Formulating the proposed model}\label{Proposed model}
In the proposed model, the system consists of three fundamental cell types forming a prey-predator-protector dynamics. Healthy or normal cells, $N$, represent the prey; tumor cells, $T$, act as the predator; and immune cells, $I$, serve as the protector. The dynamics are described by the system of equations \eqref{7}-\eqref{10}, where all parameters are assumed to be positive and constant.
For example, the coefficients $r$ and $c$ represent the growth rate of cells and the competition term among different types of cells for the limited supply of oxygen and nutrients, respectively. The definitions of other coefficients are provided in Table \ref{parameterset}.

\begin{align}
\label{7}
\dv{N(t)}{t}&=r_1 N(t) \left(1-\frac{ N(t)}{b_1}\right)- c_1 T_I(t) N(t)- c_2 T_M(t) N(t),\\
\label{8}
\dv{ T_I(t)}{ t}&=r_2 T_I(t) \left(1-\frac{ T_I(t)}{b_2}\right)+2 \alpha_I T_M(t)-\alpha_MT_I(t-\tau)-\beta_IT_I(t)-c_3  I(t)T_I(t)-c_4 N(t) T_I(t),\\
\label{9}
\dv{ T_M(t)}{ t}&=\alpha_M T_I(t-\tau)-\alpha_IT_M(t)-\beta_MT_M(t)-c_5  I(t)T_M(t)-c_6 N(t) T_M(t),\\
\label{10}
\dv{ I(t)}{ t}&=s+\rho\frac{ I (t) \left( T_I(t)+T_M(t)\right)^n}{\alpha+\left(T_I(t)+T_M(t)\right)^n}-c_7T_I(t)I(t)-c_8T_M(t)I(t)-d I(t).
\end{align}

The logistic growth of normal cells is represented by the first term of equation \eqref{7}, while the second term describes the competition between mature tumor cells and healthy cells in the host tissue.
Building on the explanation of the tumor cell life cycle in Section \ref{Depicting cell cycle by equations}, equations \eqref{8} and \eqref{9} detail tumor cell division, incorporating the delay. 

The first term of equation \eqref{8} depicts the logistic growth of tumor cells, representing the initiation of the cell cycle when new cells are born. The second term models the division of mature tumor cells into two new daughter cells during mitosis. The third term accounts for the time delay ($\tau$) required for tumor cells in interphase to mature and progress to the mitosis stage. The next term reflects the elimination of tumor cells during interphase. Finally, the last two terms capture the competition of tumor cells with the immune system cells and normal cells for nutrients and oxygen during interphase.

In equation \eqref{9}, we account for a key distinction between tumor cells in mitosis ($T_M$) and those in interphase ($T_I$). Specifically, we assume that the only source of mature tumor cells ($T_M$) is the population of tumor cells that have completed interphase and transitioned to the next phase of the cell cycle. This differs from $T_I$, where the initiation of tumor cells is represented by a logistic growth term.
Furthermore, as mitosis occurs rapidly, it is assumed to take place instantaneously, and no delay is attributed to this phase of the cell cycle.

In equation \eqref{10}, the constant $s$ represents the baseline presence of immune system cells in a healthy host, as described in \cite{kuznetsov1994nonlinear}. A healthy host refers to a state where the immune system is uncompromised and not affected by the presence of tumor cells. Lower values of  $s$ indicate a weakened immune system, suggesting insufficient immune cells under normal conditions.
The term $\rho\frac{ I (t) \left( T_I(t)+T_M(t)\right)^n}{\alpha+\left(T_I(t)+T_M(t)\right)^n}$ is a crucial component of the immune cell dynamics, incorporating parameters $\alpha$ and $ \rho$ which depend on the type of tumor cells. This term reflects the immunogenic nature of tumor cells, which stimulate an immune system response. The immunogenic properties of tumor cells trigger nonlinear growth in immune cells as they respond to the presence of tumor cells.
Mathematically, this term highlights a saturation effect in the production of immune cells, with the saturation level determined by the health of the tumor-bearing host. This saturation imposes a limit on the immune system's capacity, implying that the immune response may not always be sufficient to combat the tumor. Thus, the immune system's reaction is inherently limited in its ability to counteract the emergence of tumors effectively.

\section{Conducting an \textit{in silico} experiment on immunogenicity}\label{ex1}
\begin{figure}[htbp]
\centering
\subfloat[ Healthy host with Compromised immune system.]{\label{ns033}\includegraphics[scale=0.15]{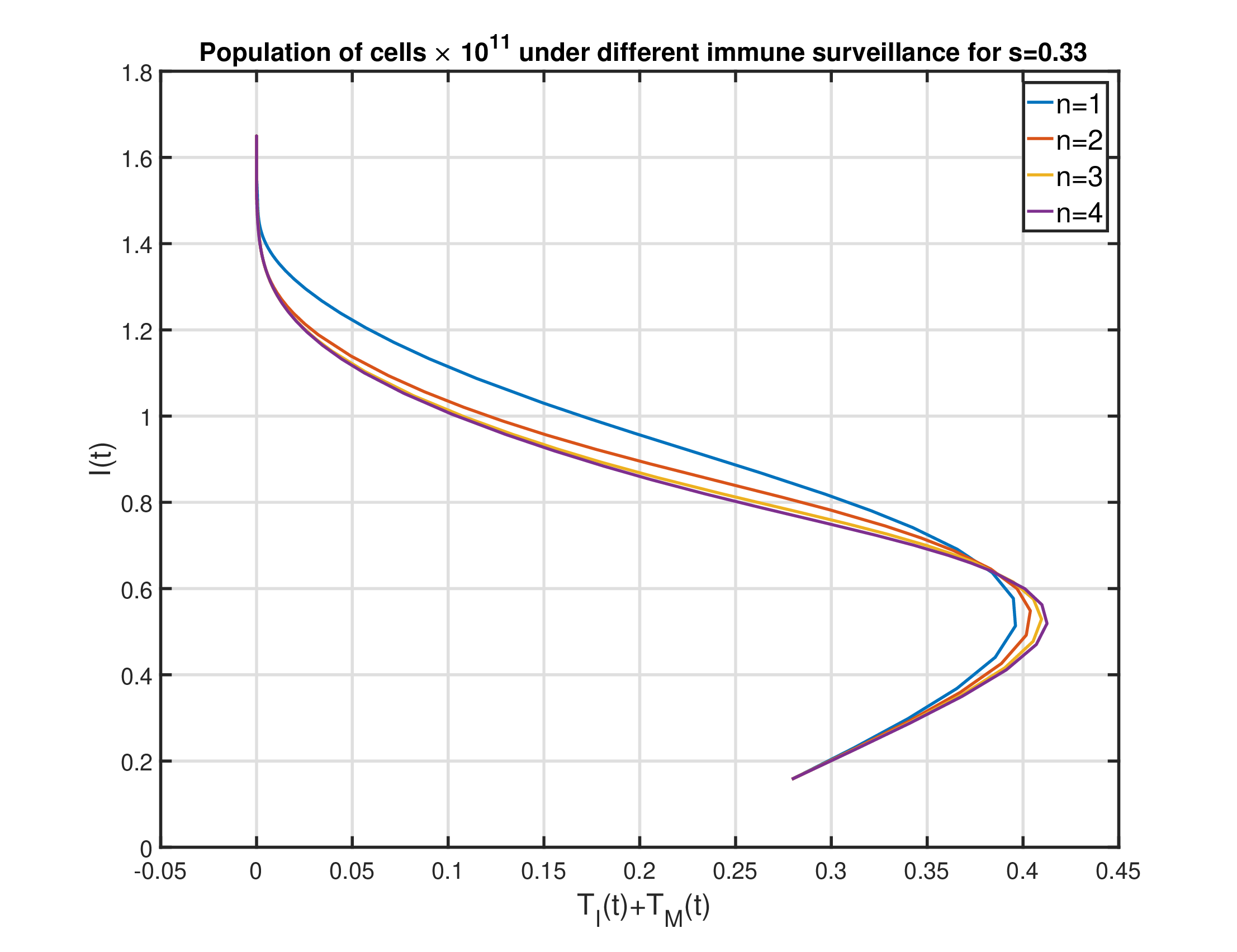}}
\subfloat[ Compromised immune system.]{\label{ns01}\includegraphics[scale=0.15]{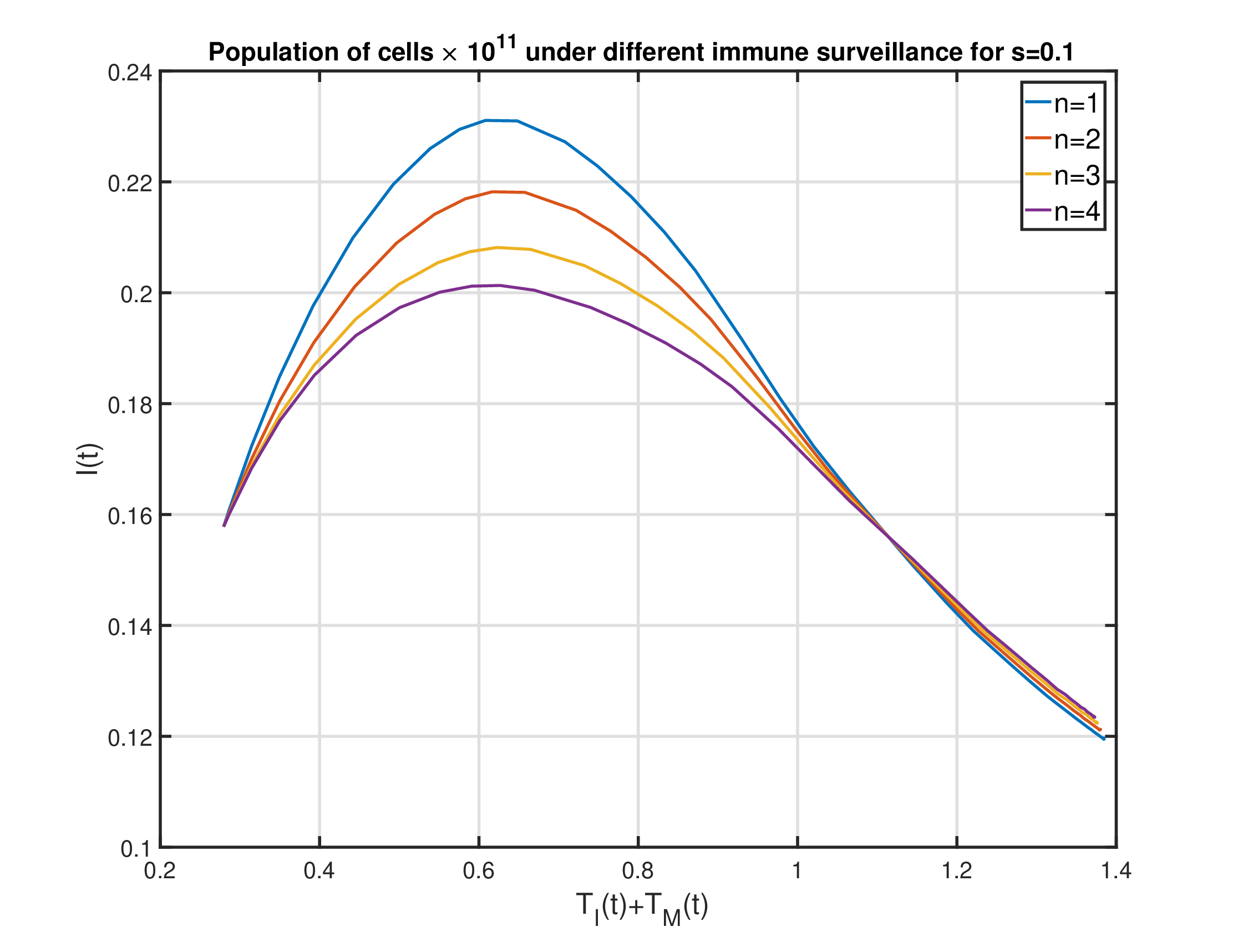}}
\caption{Immunogenicity of tumor cells for different values of $n$, given in equation \eqref{10}.
Smaller values of $n$ indicate more secretion of immune cells as a response to the mutation of tumor cells.
In a healthy host, the rise in the number of immune cells is spontaneous, unlike a compromised immune system.}
\label{NS}
\end{figure}

In this section, we examine the immunogenicity of tumor cells to gain deeper insight into the interaction between the host and tumor. As outlined in Section \ref{Proposed model}, the second term of the equation (\ref{10}) represents the immunogenic response to specific tumor cells. According to \cite{de2003dynamics}, the parameters $\alpha$ and $\rho$ define the immune threshold rate and immune response rate, respectively, while parameter $n$ represents the immune system's ability to produce immune or effector cells. Selecting an appropriate value for parameter $n$ requires a thorough understanding of its role. To provide a realistic perspective on the significance of parameter $n$, as described in equation (\ref{10}), we conduct a mathematical experiment to study the secretion of immune system cells in response to the mutation of tumor cells.
Figure \ref{NS} illustrates how the secretion of immune system cells in response to the presence of tumor cells changes with varying values of the parameter $n$.

As highlighted in Section \ref{Proposed model}, the values of certain parameters, including parameter $s$, have been adopted from existing biomathematical literature. Referring to our previous work \cite{ansarizadeh2017modelling}, we select two distinct values for the parameter $s$ to represent varying immunity levels in the tumor-bearing host. As evident from equation (\ref{10}), a higher value of the parameter $s$ corresponds to a stronger immune system.
Comparing Figures \ref{ns033} and \ref{ns01}, it is evident that a higher constant influx of immune cells significantly impacts the host's ability to respond to tumor cells. For a healthy host with $s=0.33$, the immune system promptly identifies tumor cells following their mutation, triggering the production of immune cells to combat the invaders. Consequently, immunogenicity leads to an increase in the population of immune cells. In contrast, for a compromised host with $s=0.1$, the presence of tumor cells worsens the immune system's condition, leading to a decline in immune cell numbers.

\subsection{Validation the conducted \textit{in silico} experiment}\label{ex2}
\begin{figure}[htbp]
\centering
\includegraphics[scale=0.15]{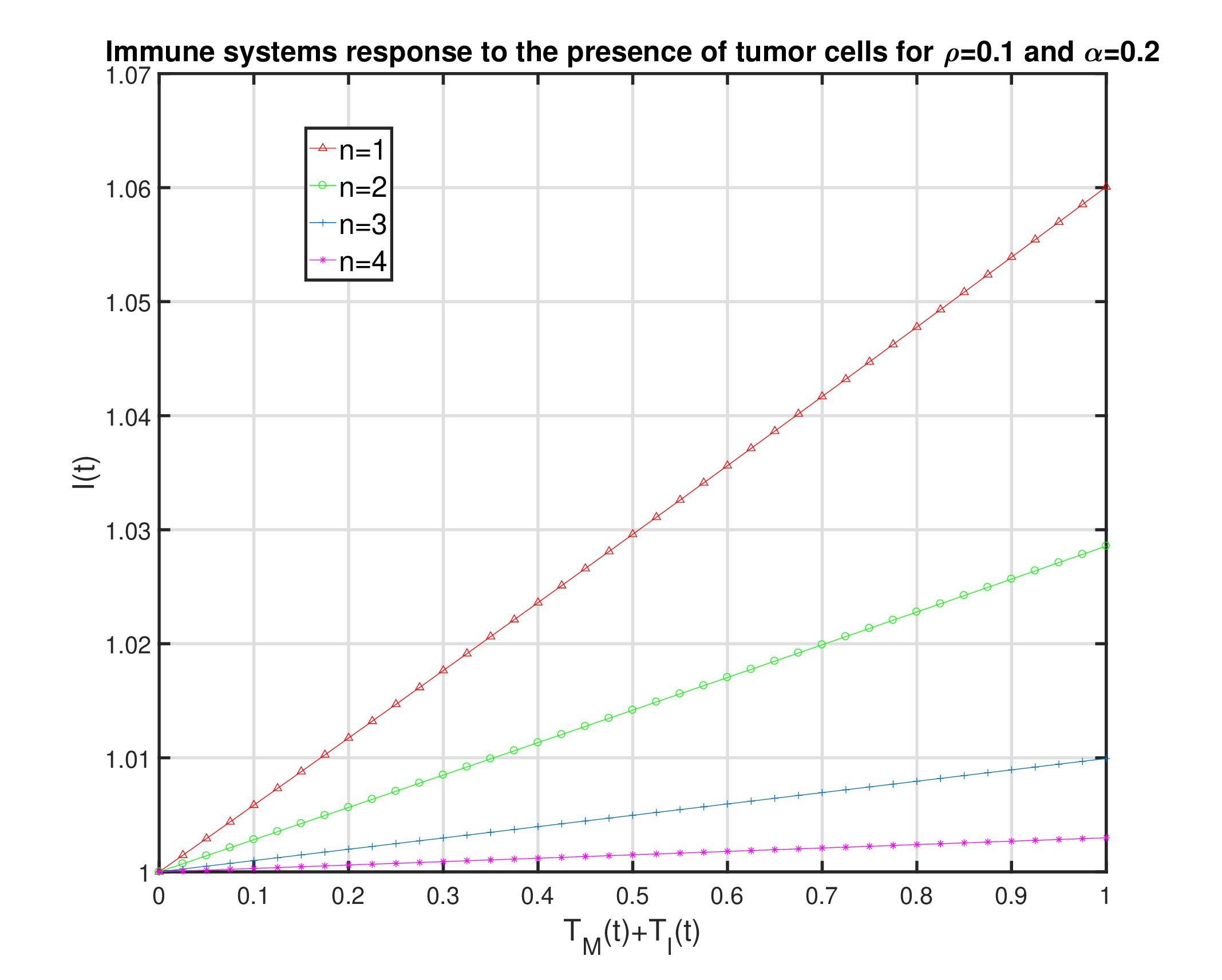}~
\includegraphics[scale=0.15]{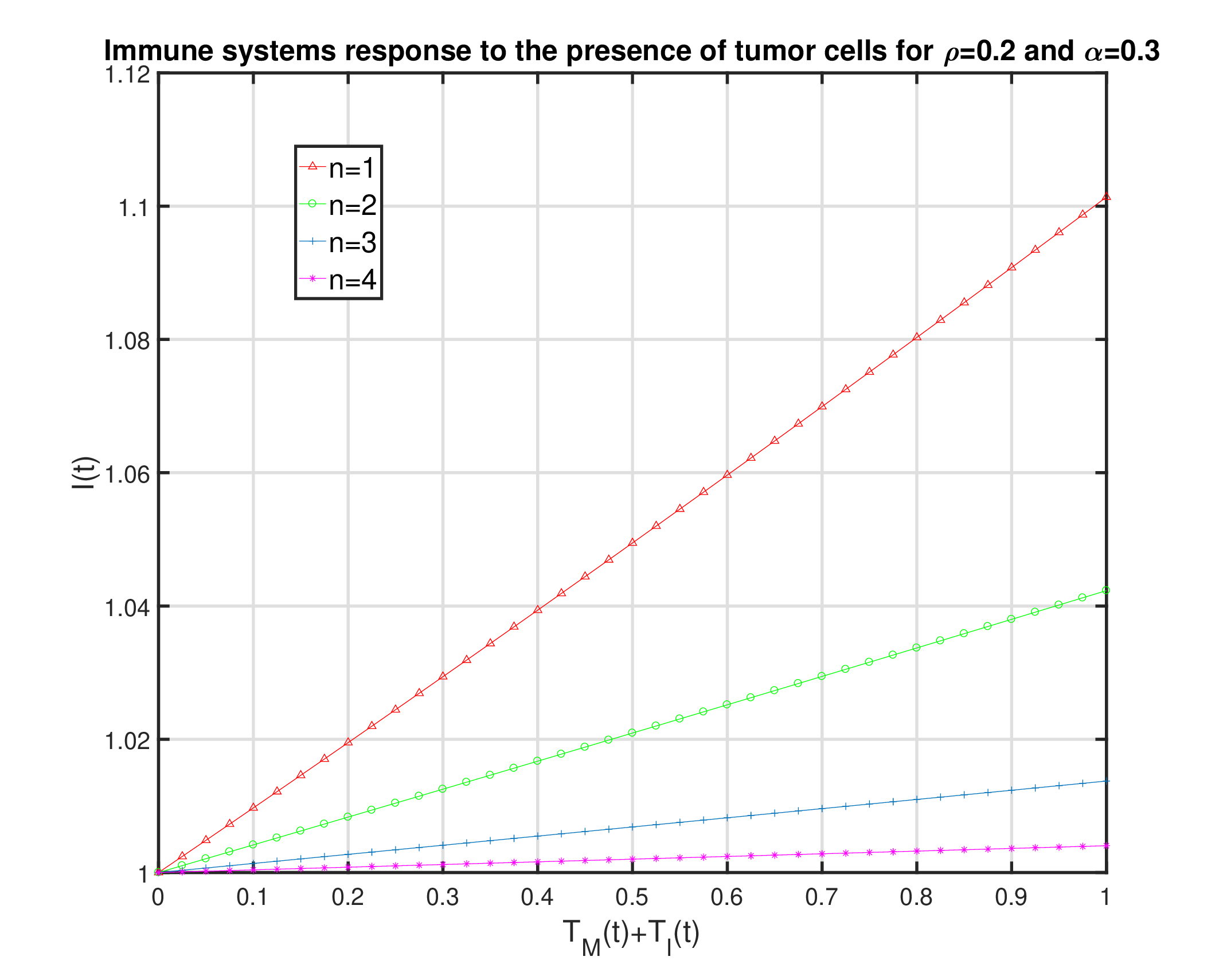}\\
\includegraphics[scale=0.15]{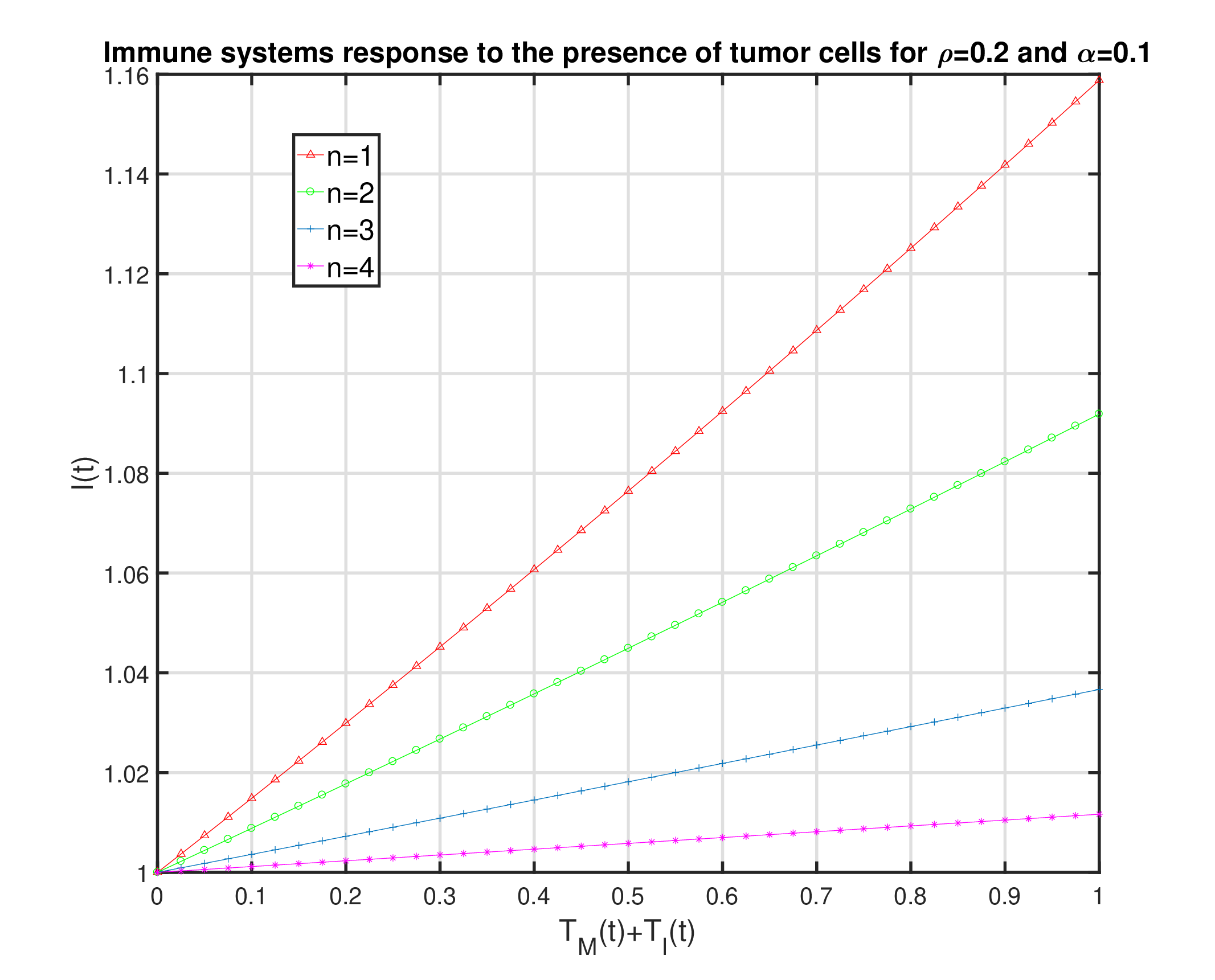}~
\includegraphics[scale=0.15]{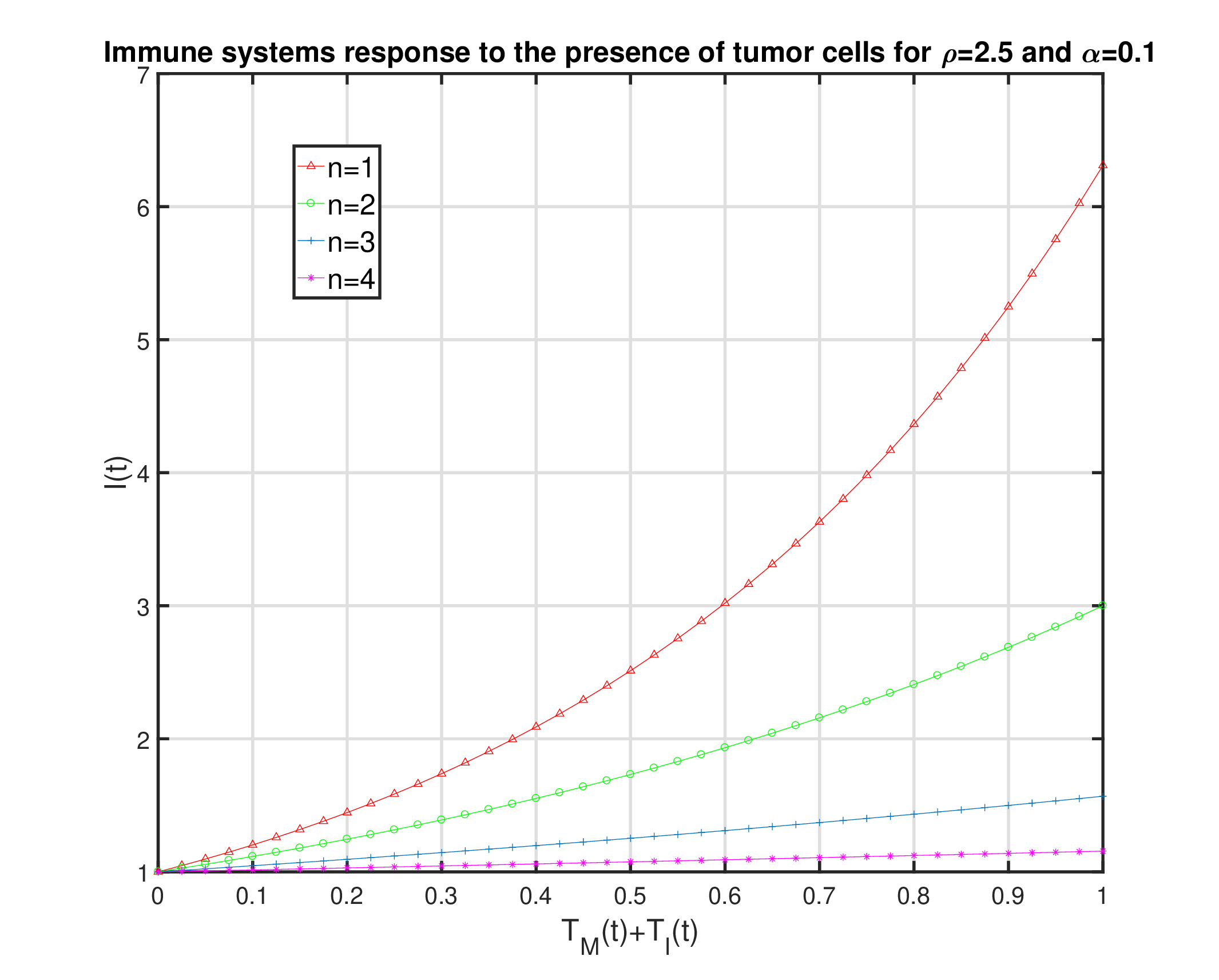}
\caption{Simulation of the first term of equation \eqref{10}, illustrating the immune system's response induced by the mutation of various tumor cell types for different values of $n$.}
\label{nsecond}
\end{figure}

To further justify the results from the \textit{in silico} experiment presented in Section \ref{ex1}, we conduct another mathematical experiment.  As shown in equation (\ref{10}), the term
$\rho\frac{ I (t) \left( T_I(t)+T_M(t)\right)^n}{\alpha+\left(T_I(t)+T_M(t)\right)^n}$
represents the nonlinear growth of immune system cells driven solely by the presence of tumor cells, a process known as immunogenicity. This term allows us to investigate how variations in the parameters related to immunogenicity affect the dynamics of the immune cell population.
The values of the parameters $\rho$, $\alpha$, and $n$ depend on the type of tumor cells and the level of immune system surveillance in the patient’s body, respectively. The increase in the population of immune system cells in response to the presence of tumor cells, as governed by these terms, is shown in Figure \ref{nsecond} for different values of $\rho$ and $\alpha$, with $n$ set to 1, 2, 3, and 4.

Figure \ref{nsecond} shows how the mutation and proliferation of tumor cells initiate the secretion of immune system cells for different values of parameters $\rho$ and $\alpha$. Regardless of the different ratios of the parameters $\rho$ to $\alpha$, smaller values of $n$ indicate a more responsive immune system.
More precisely, for a fixed population of tumor cells, and independent of the values of the parameters $\rho$ and $\alpha$, the immunogenicity term elicits a greater production of immune system cells as the value of the parameter $n$ decreases. Further mathematical experiments consistently revealed a similar trend across all ratios of $\rho$ and $\alpha$ when the value of $n$ was varied. The results of the two \textit{in silico} experiments presented in Sections \ref{ex1} and \ref{ex2} indicate that a smaller value of $n$ corresponds to a more responsive immune system, one that produces immune cells at an earlier stage of tumor formation. Furthermore, as anticipated in \cite{de2001mathematical}, for all values of $n$, the production of immune system cells induced by the antigenicity of tumor cells follows a positive, monotonically increasing, and concave trajectory.
Lastly, the influence of the parameter $n$ on the immunogenicity of tumor cells suggests that varying this parameter may offer valuable insights into how it impacts the performance of the immune system in a tumor-bearing host, both prior to and following treatment administration. Given that most studies in the literature have predominantly considered $n=1$, we adopt this value for the remainder of this work, unless explicitly stated otherwise, to ensure comparability with existing research.

\section{Parameter set for the proposed model}\label{Parameter set for the proposed model}

\subsection{Biological interpretations of the parameters}\label{Parameters}

Considering the system presented in \eqref{15}-\eqref{18} and the established certainties discussed herein, an effort is made to identify and select a suitable set of parameters for the proposed model.

\begin{description}
\item[$\bullet$] 
Histological experiments indicate that tumor cell malignancy in humans is higher during the $S$  phase \cite{matthews2022cell}.
Theoretically, tumor cells in interphase, $T_I$, are assumed to exhibit greater aggressiveness and a higher capacity to eliminate normal and immune system cells compared to tumor cells in mitosis, $T_M$.
Consequently, $T_I$ cells are hypothesized to exhibit greater resistance to immune system cells. This biological observation is mathematically represented by the following inequalities: $c_1>c_2$,~ $c_7>c_8$,~ $c_6>c_4$.

\item[$\bullet$] 
As noted in \cite{hanahan2000hallmarks}, tumor cells, unlike normal cells, exhibit minimal dependency on external growth signals due to their self-sufficiency in generating these signals. Consequently, the following choice is considered reasonable: $r>1$.

\item[$\bullet$] Given that tumor cells aggressively compete with healthy host tissue cells for nutrients, oxygen, and an acidic environment \cite{paul2022tumor}, we assume the following: $c_1>c_4$,~$ c_2>c_6$.

\item[$\bullet$] Alternatively, experiments and clinical observations in \cite{kuznetsov1994nonlinear} have proposed acceptable ranges for other parameters, including $s$, $\rho$, and $\alpha$, as ~$0 \leq s \leq0.5$, ~ $0<\rho<2.5$, ~$b_2\leq b_1\leq 1$.

\item[$\bullet$] Immune system cells are expected to target cancer cells, not normal cells, and the rate of tumor cell lysis by normal cells is assumed to be minimal if not zero. Therefore, the relationships between the relevant parameters are: $c_3>c_4$, $c_5>c_6$.

\end{description}

All the aforementioned inequalities are based on our assumptions derived from the relevant literature, which inform the customization of each parameter set.

\subsection{Nondimensionalisation of the parameters}\label{Nondimensionalisation of the parameters}
By defining the nondimensional variables as $\hat N=\frac {N}{N_0}$, $\hat T_I=\frac {T_I}{T_{I_0}}$, $\hat T_M=\frac {T_I}{T_{M_0}}$, $\hat t=t r_1$, and substituting them into the given system of equations while neglecting the delay, the nondimensionalized parameters can be determined. The resulting equation of \eqref{7} is:
\begin{equation*}
\label{9}
\dv{\hat N(t)}{\hat t}=\hat N(t) \left(1-\frac{ \hat N(t)}{\frac{b_1}{N_0}}\right)- \dfrac{c_1 T_{I_0}}{r_1}  ~\hat T_I(t) \hat N(t)- \dfrac{c_2 T_{M_0}}{r_1}  ~\hat T_M(t) \hat N(t).
\end{equation*}
Denote $\hat b_1 = \dfrac{b_1}{N_0}$, $ \hat c_1 = \dfrac{c_1T_{I_0}}{r_1}$, and  $\hat c_2 = \dfrac{c_2T_{M_0}}{r_1}$. Then, after dropping \^~, the normalized form of equation \eqref{7} reads:
\begin{equation*}
\label{9}
\dv{ N(t)}{ t}= N(t) \left(1-\frac{N(t)}{b_1}\right)- c_1 T_I(t)  N(t)- c_2 T_M(t)  N(t).
\end{equation*}

Repeating this procedure for the other equations yields the following nondimensionalized variables:
\\
\vspace{-5mm}
\begin{itemize}
\item[$\bullet$] From equation \eqref {8}:
\\
$r = \dfrac{r_2 }{r_1 }$,
$\hat b_2 = \dfrac{b_2 }{T_{I_0} }$,
$\hat \alpha_I = \dfrac{\alpha_I }{r_1 }$,
$\hat \alpha_M = \dfrac{\alpha_M}{r_1 }$,
$\hat \beta_I =\dfrac{\beta_I  }{r_1 }$,
$\hat c_3 = \dfrac{c_3 I_0}{r_1}$,
$\hat c_4 = \dfrac{c_4 N_0}{r_1}$.
\item[$\bullet$] From equation \eqref {9}:
\\
$\hat \alpha_M = \dfrac{\alpha_M }{r_1 }$,
$\hat \alpha_I = \dfrac{\alpha_I }{r_1}$,
$\hat \beta_M = \dfrac{\beta_M  }{r_1 }$,
$\hat c_5 = \dfrac{c_5 I_0}{r_1 }$,
$\hat c_6 = \dfrac{c_6 N_0}{r_1}$.

\item[$\bullet$] 
It is assumed that immune system cells have a greater impact on tumor cells during mitosis, as tumor cells are more malignant in interphase \cite{awang2022tumor}. As a result, it is reasonable to set $c_5$ larger than $c_6$.  Hence, from equation \eqref{10} it is conclude:
$\hat s = \dfrac{s}{r_1 I_0}$,
$\hat \rho = \dfrac{\rho}{r_1} $,
$\hat \alpha = \dfrac{\alpha}{T_{M_0}}$,  
$\hat c_7 =  \dfrac{c_7 T_{I_0}}{r_1}$, 
$\hat c_8 = \dfrac{c_8 T_{M_0}}{r_1}$,  
$\hat d= \dfrac{d}{r_1}$.
\end{itemize}
Finally, we arrive the normalized system of equations:
\begin{align}
\label{15}
&\dot  N(t) =  N(t) \left(1-\frac{ N(t)}{b_1}\right)- c_1T_I(t) N(t)- c_2 T_M(t) N(t),\\
\label{16}
&\dot T_I(t) =  r T_I(t) \left(1-\frac{ T_I(t)}{b_2}\right)+2\alpha_IT_M(t)-\alpha_MT_I(t-\tau)-\beta_IT_I(t) -c_3  I(t)T_I(t)-c_4 N(t) T_I(t), \\ 
\label{17}
&\dot T_M(t) =  \alpha_MT_I(t-\tau)-\alpha_IT_M(t)-\beta_MT_M(t)-c_5 I(t) T_M(t) -c_6N(t) T_M(t),\\
\label{18}
&\dot I(t) =  s+\rho\frac{ I (t) \left( T_I(t)+T_M(t)\right)}{\alpha+T_I(t)+T_M(t)}-c_7 T_I(t)I(t)-c_8T_M(t)I(t)-d I(t).
\end{align}
\begin{table*}[htpb]
\caption{Nondimensionalized set of parameters}
\centering
\begin{threeparttable}
\begin{tabular} {c l  c c }\Xhline{1.2pt}
\thead {Parameter}&\thead {Description}&\thead {Value}&\thead {Source}\\ \Xhline{1.2pt}
{\small$b_1$}&{\small Normal carrying capacity}&{\small \makecell[l] {$1$}}&\cite{ansarizadeh2017modelling}\\ \hline
{\small$c_1$}&{\small \makecell[l] {Competition of tumor cells  \\with normal cells}}&{\small \makecell[l] {$0.948$}}&\tnote{*}\\ \hline
{\small$c_2$}&{\small \makecell[l] {Competition of tumor cells  \\with normal cells}}&{\small \makecell[l] {$0.9$}}&\tnote{*}\\ \hline
{\small $r$}&{\small  {tumor cells growth rate} }&{\small \makecell[l]{$1.1$}}&\cite{ansarizadeh2017modelling}\\ \hline
{\small$b_2$}&{\small tumor carrying capacity}&{\small \makecell[l] {$0.81$}}&\cite{ansarizadeh2017modelling}\\ \hline
{\small $\alpha_I$}&{\small\makecell[l] {Fraction of  cells cycling\\from mitosis to interphase}  }&{\small \makecell[l]{$0.8$}}&\cite{newbury2007numerical}\\ \hline
{\small $\alpha_M$}&{\small\makecell[l] {Fraction of  cells cycling\\from interphase to mitosis}  }&{\small \makecell[l]{ $0.98$}}&\cite{newbury2007numerical}\\ \hline
{\small$\beta_I$}&{\small \makecell [l] {Death rate\\in interphase}}&{\small \makecell[l] {$0.11$}}&\cite{newbury2007numerical}\\ \hline
{\small$c_3$}&{\small \makecell [l]{ Rate of tumor cell inactivation \\by immune cells}} &{\small \makecell[l]{$1$}}&\tnote{*}\\ \hline
{\small$c_4$}&{\small \makecell [l]{ Rate of tumor cell inactivation \\by immune cells}} &{\small \makecell[l]{$0.015$}}&\tnote{*}\\ \hline
{\small$\beta_M$}& {\small \makecell [l]{Death rate \\in mitosis}}&{\small \makecell[l] {$0.4$}}&\cite{newbury2007numerical}\\ \hline
{\small$c_5$} &{\small \makecell[l]{Rate of tumor cell inactivation \\by immune cells}}&{\small \makecell[l] {$1$}} &\tnote{*}\\ \hline
{\small$c_6$}&{\small \makecell[l] {Competition of normal cells \\with tumor cells}}&{\small \makecell[l]{ $0.065$}}&*\\ \hline
{\small$s$}&{\small \makecell[l]{Steady inflow of\\ immune system cells}}&{\small \makecell[l] {$0.33 $}}&\cite{ansarizadeh2017modelling}\\ \hline
{\small$\rho$}&{\small \makecell[l]{Immune system\\ response rate}}&{\small \makecell[l] {$0.2$}}&\cite{ansarizadeh2017modelling}\\ \hline
{\small$\alpha$}&{\small \makecell[l]{Threshold of immune system\\trigered by tumor cells}}&{\small \makecell[l] {$0.3$}}&\cite{ansarizadeh2017modelling}\\ \hline
{\small $c_7$}&{\small Immune tumor competition}&{\small \makecell[l] {$0.6$}}&\tnote{*}\\ \hline
{\small $c_8$}&{\small Immune tumor competition}&{\small \makecell[l] {$0.55$}}&\tnote{*}\\ \hline
{\small$d$}&{\small Death rate of immune cells}&{\small \makecell[l]{$0.2$}}&\cite{ansarizadeh2017modelling}\\ \hline \Xhline{1.2pt}
\end{tabular}
\begin{tablenotes}
\item[*] Competition terms have been chosen in a valid range $ 0 \leq c_i \leq 1$
\end{tablenotes}
\end{threeparttable}
\label{parameterset}
\end{table*}

\section{Model analysis}\label{Model analysis}
In this section, the equilibrium points and their stability are investigated.

\subsection{Equilibrium points}\label{Equilibrium points}

Set the left-hand sides of the system \eqref{15}-\eqref{18} to zero and we obtain:
\begin{align}
\label{pe1}
&0  =  N \left(1-\frac{ N}{b_1}\right)- c_1T_I N- c_2 T_M N,\\
\label{pe2}
 &0 =  r T_I \left(1-\frac{ T_I}{b_2}\right)+2 \alpha_I T_M-\alpha_MT_I-\beta_IT_I-c_3  IT_I-c_4 NT_I,\\
\label{pe3}
&0  =  \alpha_MT_I-\alpha_IT_M-\beta_MT_M-c_5 I T_M -c_6 N T_M,\\
\label{pe4}
&0  =  s+\rho\frac{ I \left( T_I+T_M\right)}{\alpha+T_I+T_M}-c_7 T_II-c_8T_MI-d I.
\end{align}
Based on system \eqref{pe1}-\eqref{pe4}, we will examine the eixstence of equilibrium points and their stability. We begin with the tumor-free case, followed by an analysis of the coexistence scenario.

\subsubsection{Equilibrium points in tumor-free case}\label{Equilibrium points in tumor-free case}

In this case, there are no tumor cells in the host, implying $T_I=0$ and $T_M=0$. Therefore, the system of algebraic equations \eqref{pe1}-\eqref{pe4} reduces to:
\begin{align}
&N^o \left(1-\frac{ N^o}{b_1}\right)=0,\\
&s-d I^o=0.
\end{align}
By substituting the parameters provided in Table \ref{parameterset}, the equilibrium point in the tumor-free situation is:
\begin{align}
P^o(b_1,~ 0 ,~ 0, ~\dfrac{s}{d})=(1,~0,~0,~1.65).
\end{align}
The assumption of $T_I=0$ and $T_M=0$ significantly simplifies the process of determining $P^o$.

\subsubsection{Equilibrium points in coexisting case}\label{Points of equilibria in coexisting case}


It is important to note that the introduction of delay into the system does not affect the equilibrium points. 
Numerical methods indicate the presence of an equilibrium point in the coexistence case for the proposed system and the specified parameter set, as illustrated in Table \ref{PE}. 
\begin{table}[htp]
\caption{Equilibrium points for tumor-free and coexistence case for the system of equations (\ref{pe1})-(\ref{pe4})}
\centering
\begin{tabular} {c|c}\Xhline{1.2pt}
\thead {tumor-free case}&\thead {Coexistence case}\\ \Xhline{1.2pt}
$P^o(N^o, T_I^o, T_M^o, I^o)$&$P^*(N^*, T_I^*, T_M^*, I^*)$\\
(1, 0, 0, 1.65)&(0.4731, 0.3499, 0.2125, 0.6439)\\ \Xhline{1.2pt}
\end{tabular}
\label{PE}
\end{table}

However, determining whether the equilibrium point identified through numerical methods is unique remains a contentious issue. In this study, the existence and uniqueness of equilibrium points have been verified using MATCONT.
With all prerequisites in place, the stability of the proposed system under the given parameter set can now be examined.

In the following sections, the Jacobian matrix of equations (\ref{15})-(\ref{18}) and its characteristic equation are used with the Routh-Hurwitz criteria to assess the stability of equilibrium points in tumor-free and coexistence scenarios.




\subsection{Local stability in tumor-free case}\label{Stability in tumor-free case}

The time delay should be excluded in the tumor-free scenario since the discrete delay in the system \eqref{15}-\eqref{18} represents the intrinsic delay in the life cycle of tumor cells. Under this condition, the proposed system of delay differential equations \eqref{15}-\eqref{18} reduces to the system of ordinary differential equations \eqref{115}-\eqref{118}:
\begin{align}
\label{115}
&\dot  N(t)= N(t) \left(1-\frac{ N(t)}{b_1}\right)- c_1T_I(t) N(t)- c_2 T_M(t) N(t),\\
\label{116}
&\dot T_I(t)=r T_I(t) \left(1-\frac{ T_I(t)}{b_2}\right)+2 \alpha_I T_M(t)-\beta_IT_I(t)-c_3  I(t)T_I(t)-c_4 N(t) T_I(t),\\
\label{117}
&\dot T_M(t)=-\alpha_IT_M(t)-\beta_MT_M(t)-c_5 I(t) T_M(t) -c_6 N(t) T_M(t),\\
\label{118}
&\dot I(t)=s+\rho\frac{ I (t) \left( T_I(t)+T_M(t)\right)}{\alpha+T_I(t)+T_M(t)}-c_7 T_I(t)I(t)-c_8T_M(t)I(t)-d I(t).
\end{align}
For this system, calculating the Jacobian matrix and characteristic equation is straightforward.
\begin{align*}
{\boldsymbol J}({\hat N}, {\hat T_I} ,{\hat T_M}, {\hat I})=\left[\begin{array}{cccc}
a_{11}& -c_1 {\hat N} & -c_2 {\hat N}&0\\
-c_4 {\hat T_I}& a_{22}&2\alpha_I & -c_3 {\hat T_I}\\
-c_6 {\hat T_M} &0 &a_{33} &-c_5 {\hat T_M} \\
 0&a_{42} &a_{43} &a_{44}
\end{array}\right],
\end{align*}
where
\begin{align*}
&a_{11}= 1-2\dfrac{ {\hat N} }{b_1}-c_1 {\hat T_I} -c_2  {\hat T_M},\\
&a_{22}=r(1-2\dfrac{  {\hat T_I} }{b_2})-\beta_I-c_3  {\hat I} -c_4 {\hat N},\\
&a_{33}=-\alpha_I-\beta_M-c_5 {\hat I} -c_6 {\hat N},\\
&a_{42}=\dfrac{\alpha \rho {\hat I} }{(\alpha+ {\hat T_I}+{\hat T_M})^2}-c_7 {\hat I},\\
&a_{43}=\dfrac{\rho \alpha  {\hat I}}{(\alpha+ {\hat T_I}+{\hat T_M})^2}-c_8 {\hat I},\\
&a_{44}=\rho\dfrac{{\hat T_I}+{\hat T_M}}{\alpha+{\hat T_I}+{\hat T_M}}-c_7 {\hat T_I}-c_8 {\hat T_M}-d.
\end{align*}
The general form of the characteristic equation of the Jacobian matrix in the tumor-free case is $\lambda ^ 4+ P_1 \lambda ^3+ P_2 \lambda ^2+ P_3 \lambda + P_4=0$.
Upon substitution of the parameters, the characteristic equation becomes
\begin{align}
\lambda ^4 + 4.79 \lambda^3 +  6.47563  \lambda^2 +  3.07915 \lambda+0.393525=0.
\label{cetf}
\end{align}
\begin{table}[htp]
\caption{Stability of equilibrium point in tumor-free case}
\centering
\begin{tabular} {c|c c c c}\Xhline{1.2pt}
\diagbox[width=10em]{\thead {Point of \\equilibrium}}{\thead {RH \\criteria}}&\thead {$P_1>0$}&\thead {$P_3>0$}&\thead {$P_4>0$}&\thead {$P_1P_2P_3>P^{2}_3+P^{2}_1P_4$}\\ \Xhline{1.2pt}
$P^o$&$4.79$&$3.07915$&$0.393525$&$95.5098>18.5102$\\ \Xhline{1.2pt}
\end{tabular}
\label{RHTF}
\end{table}

As shown in Tables \ref{RHTF} and \ref{eigenvaluesTF}, all Routh-Hurwitz conditions are satisfied so all eigenvalues of the system are negative. Therefore, the proposed model in the tumor-free case, with the parameter set provided in Table \ref{parameterset}, is stable.
\begin{table}[htp]
\caption{Eigenvalues of the equation(\ref{cetf})}
\centering
\begin{tabular} {c|c c c c}\Xhline{1.2pt}
\diagbox[width=10em]{\thead {Point of \\equilibrium}}{\thead {Eigenvalues}}&\thead {$\lambda_1$}&\thead {$\lambda_2$}&\thead {$\lambda_3$}&\thead {$\lambda_4$}\\ \Xhline{1.2pt}
$P^o$&$-2.915 $& $-1$&$-0.675$&$-0.2$ \\ \Xhline{1.2pt}
\end{tabular}
\label{eigenvaluesTF}
\end{table}

\subsection{Stability of the system in coexistence case}\label{Stability of the system in coexistence case}


To illustrate how the incorporation of delay results in more realistic outcomes when analyzing these interactions, the coexistence scenario is examined in both the absence and presence of delay.

\subsubsection{Coexistence case without delay}


Similar to the tumor-free case, the Routh-Hurwitz criteria are applied to investigate the stability of the system under these conditions. By substituting the parameters and the equilibrium point, the characteristic equation becomes:
\begin{align}
\lambda^4 + 3.35243 \lambda^3 + 3.29579 \lambda^2 + 0.976702 \lambda + 0.0311475=0.
\label{cewtnd}
\end{align}
The stability of the system is then verified in Tables \ref{RHWTND} and \ref{eigenvaluesWTND}.
\begin{table}[htp]
\caption{Stability considerations for equilibrium point, $P^*$, in absence of delay}
\centering
\begin{tabular} {c|c c c c}\Xhline{1.2pt}
\diagbox[width=10em]{\thead {Point of \\equilibrium}}{\thead {RH \\criteria}}&\thead {$P_1>0$}&\thead {$P_3>0$}&\thead {$P_4>0$}&\thead {$P_1P_2P_3>P^{2}_3+P^{2}_1P_4$}\\ \Xhline{1.2pt}
$P^*$&$3.35243$&$0.9767$&$0.031147$&$10.7915>1.30401$\\ \Xhline{1.2pt}
\end{tabular}
\label{RHWTND}
\end{table}


\begin{table}[htp]
\caption{Eigenvalues of the equation (\ref{cewtnd})}
\centering
\begin{tabular} {c|c c c c}\Xhline{1.2pt}
\diagbox[width=10em]{\thead {Point of \\equilibrium}}{\thead {Eigenvalues}}&\thead {$\lambda_1$}&\thead {$\lambda_2$}&\thead {$\lambda_3$}&\thead {$\lambda_4$}\\ \Xhline{1.2pt}
$P^*$&$-1.854$& $-0.995$&$-0.467$&$ -0.0365$\\ \Xhline{1.2pt}
\end{tabular}
\label{eigenvaluesWTND}
\end{table}

\subsubsection{Coexistence case with delay}
The average cell cycle duration in human solid tumors is approximately 2 days. Given that the $G_1$ and $M$ phases have the longest and shortest durations, respectively, we assume that the 2-day duration pertains to interphase \cite{tubiana1976comparison}. This interphase duration will serve as the time delay utilized in our subsequent numerical analysis.

Re-write the system  \eqref{15}-\eqref{18} in matrix form as $\dot  X(t)=\boldsymbol F(t, \boldsymbol X(t), \boldsymbol X(t-\tau))$. Then,
the linearisation of the system around the equilibrium point leads to $\dot X(t)=\boldsymbol {J} \boldsymbol X(t)+ \boldsymbol J_{\tau} \boldsymbol X(t-\tau)$
where, $\boldsymbol J_{ \it i\it j}=\frac{\partial \boldsymbol F_{\it i } }{\partial \boldsymbol X_{\it j}}\vert_{i,j=1,2,3,4}$
is the same as Jacobian matrix given in Section \ref{Stability in tumor-free case} and
\begin{align}
{\boldsymbol J_{\tau}}=\left[\begin{array}{c c c c}
0 & 0& 0 &0 \\
 0& -\alpha_M &0& 0\\
 0 &\alpha_M &0& 0\\
 0 &0 &0& 0
\end{array}\right].
\end{align}
Therefore, the characteristic equation is:
\begin{align}
 {\rm det} \boldsymbol M (\lambda)= |\lambda \boldsymbol I-\boldsymbol J- \exp ^{-\lambda \tau}\boldsymbol J_{\tau}| = \boldsymbol D (\lambda (\tau), \tau)=P(\lambda)+e^{-\lambda \tau} Q(\lambda)  =0.
\label{c.e.}
\end{align}
In the case of $(\tau \neq 0)$, only the coexistence scenario should be considered, as $\tau$ represents the intrinsic delay in the  life cycle of tumor cells. By substituting the parameters and the equilibrium point of the coexistence case, the characteristic equation \eqref{c.e.} becomes
\begin{align}
{\rm det} \boldsymbol M (\lambda)&= \boldsymbol D (\lambda (\tau), \tau)=\lambda^4 + 3.35 \lambda^3 + 3.29 \lambda^2 + 0.976  \lambda+0.031 
\nonumber 
\\
&+ e^{-\lambda \tau} (0.98  \lambda^3 +1.12  \lambda^2 + 0.241 \lambda-0.031)=0.
\label{characequ}
\end{align}
Increasing $\tau$ may cause the roots to move across the complex plane, potentially leading to stability switching that happens when \eqref{characequ} has a purely imaginary root, $\lambda=i \omega$, where $\omega>0$. Thus,
$D(i \omega, \tau)=P(i \omega)+Q(i \omega) e^ {-i \omega \tau}=0$.
After substituting the numeric values of the parameters, the expressions for $P (i \omega)$ and $Q (i \omega)$ are as follows:
\begin{align*}
P(i \omega)&= \omega^4 - 3.294 \omega ^2 + 0.0313867 +  i (0.975773 \omega  -  3.35154  \omega^3).
\\
Q(i \omega)&=- 1.11739 \omega^2  - 0.03124+  i ( 0.240705  \omega  - 0.98  \omega^3),
\end{align*}
and define a function $F(\omega)$:
\begin{align}
\label{F}
F(\omega)=|P(i \omega)|^2-|Q(i \omega)|^2 =0.
\end{align}
Solving equation \eqref{F} for $\omega>0$ reveals whether increasing $\tau$ induces instability in the system. To explore this, the numerical values are substituted into $F(\omega)$, yielding:
\begin{align}
\label{replaced}
F(\omega)&=\omega^8+ 3.6838 \omega^6+ 3.59773 \omega^4+ 0.617586  \omega^2 +9.23063 \times 10^{-6}=0
\end{align}
that has the roots:
\begin{align}
\omega= ~\pm  0.00386622 i,~\pm   0.465875 i,~\pm  1.1508 i, ~\pm 1.46032 i.
\end{align}
This analysis demonstrates that the system does not exhibit stability switching for a delay of 2 days. 
Given the biological basis for the selected delay value, our system does not encounter prolonged delays, thereby maintaining stability.

\section{History function}

One of the novelties of this work, aligned with biological interpretations, is the assignment of ascending history functions for tumor cells across different phases of the cell cycle. Given that early population growth is characterized as either exponential, Gompertzian, or approximately linear in logistic models, a monotonically increasing function is a logical choice for describing tumor cell evolution during early formation. This approach offers a more reasonable history function, as the exact one remains unknown \cite{villasana2004heuristic}.

\begin{figure}[htpb]
\centering
\includegraphics[width=0.5\textwidth]{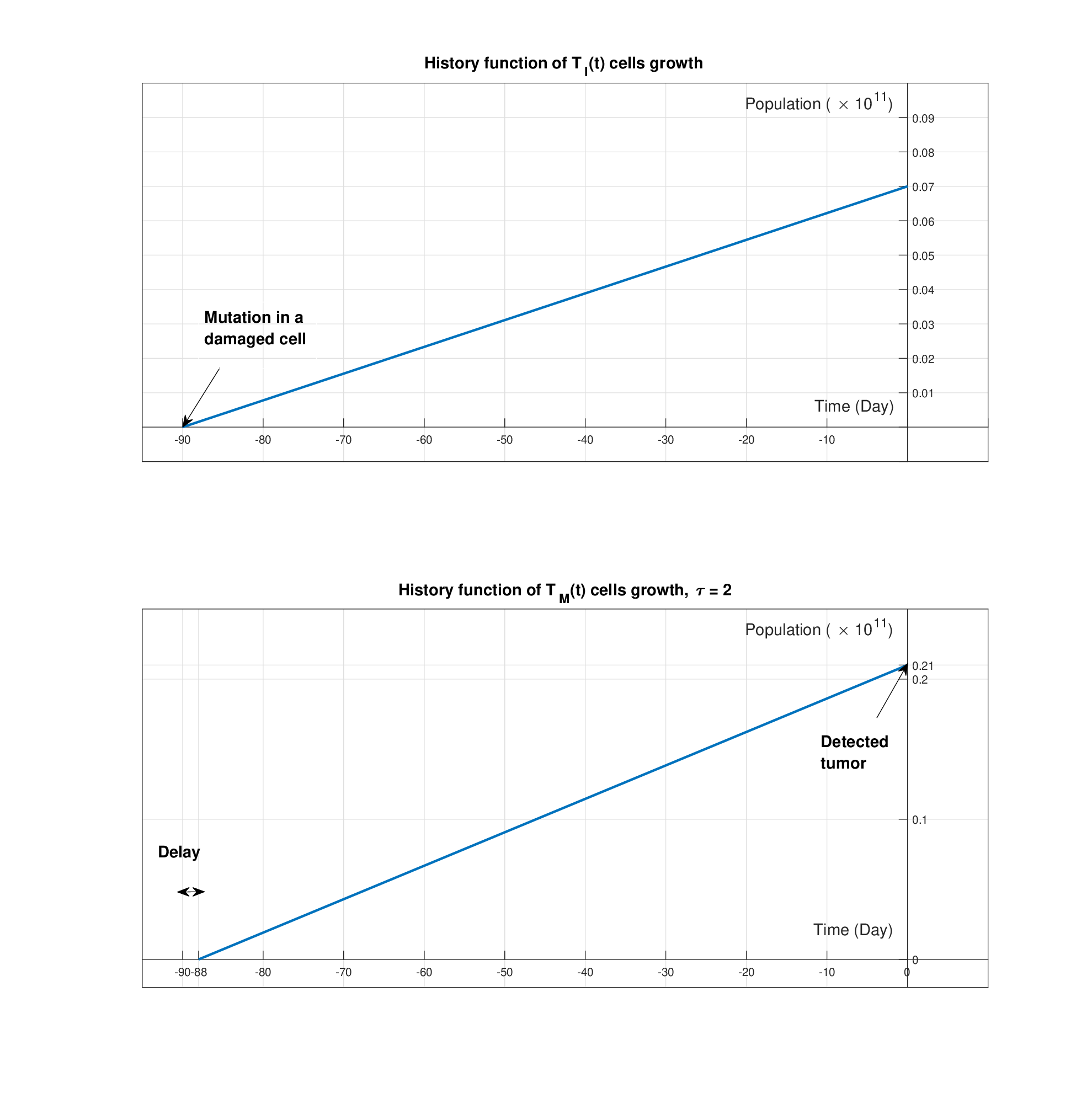}
\caption{
Ascending growth of tumor cells before detection. The mutation is assumed to occur 90 days before detection, with a 2-day delay for tumor cells transitioning from $T_I$ to $T_M$ cells, resulting in the initiation of $T_M$ cell growth 88 days before detection. The slope of the history function is steeper for $T_I$ cells due to their increased malignancy.}
\label{historyfunction}
\end{figure}

The formulation of history functions in this work is biologically motivated. Since tumor cell populations are split into $T_I$ and $T_M$, representing cells at different cell cycle stages, each group requires its corresponding history function. These separate functions illustrate cell evolution before the tumor manifests.
Secondly, recognizing that mitosis occurs instantly relative to interphase and assuming each cycle begins from interphase, we propose that for the first cycle starting at  $t_0$, mitosis occurs at $t_0+\tau$, with $t_1=t_0+\tau$ marking the start of the next cycle. A period function is defined later in this section to mathematically represent this repetitive cell division process.
Considering the dormancy period, $t_0$ represents the initiation of mutation, occurring well before tumor manifestation. From a mathematical perspective \cite{wu2010sensitivity}, the history function must cover at least $\tau$ before the phenomenon begins. In Figure \ref{historyfunction}, we assume that mutation starts three months before tumor detection. Given our assumption that the cell cycle begins at interphase and takes approximately two days for tumor cells to undergo mitosis $\tau=2 $ days, the history function for $T_M$ cells starts two days after that of $T_I$ cells. Both populations increase at prescribed rates until tumor detection.
The analysis of the quantitative patterns of history functions relies on the initial populations of the cell types. To simplify the model, the tumor is assumed to be in its early stages, excluding complexities like heterogeneity, angiogenesis, and mitosis. Breast tumors under $2 cm$ are considered early-stage, and for this study, the tumor size is assumed to be $1 cm$.

Assuming the normalized initial distribution of tumor cells from our previous work \cite{ansarizadeh2017modelling}, spread over the tumor site $-0.5~cm<x<0.5~cm$, the following functions represent the initial populations of different cell types across the tumor site.
\[
 \begin{cases}
 \ T(x,0)=1-0.75 \rm Sech(x), & \textit{}~~~~~-0.5<x<0.5, \\
 \ N(x,0)=0.2 ~\rm exp(-2x^2), & \textit{}~~~~~-0.5<x<0.5,\\
 \ I(x,0)=0.375-0.235 \rm Sech^2(x), & \textit{}~~~~~-0.5<x<0.5, \\
 \end{cases}
 \label{his}
 \]
and the initial population of tumor cells will be:
$$\int _{x_{left}}^{x_{right}} T(x)\;\mathrm{d}x=\int _{-0.5}^{0.5} 1-0.75 \rm Sech(x)\;\mathrm{d}x=0.27942838129940584\approx0.28 $$
Since the initial distribution function is normalized, the real population of cells is obtained by multiplying the function by $10^{11}$.
Therefore, the initial population of tumor cells is $2.8 \times 10^{10}.$
This population is supposed to be the total number of tumor cells including those in interphase and mitosis ($T_I+T_M$).
The threshold for tumor cell detection by clinical equipment is approximately $10^7$  cells \cite{shochat1999using}. Under the given assumptions, the tumor cell population exceeds the detection threshold. However, it is important to note that tumors are typically not treated until they are detected. Unfortunately, most patients delay action against the signs and symptoms of the tumor. To represent a more realistic scenario, the tumor cell population at the point of detection is assumed to be significantly larger than what can initially be detected. Additionally, the exact number of cells is not the primary concern; rather, the proportion of cells plays a critical role in determining the outcome of interactions among different cell types.

Based on the cell cycle depicted in Figure \ref{cellcycle}, reasonable assumptions are made to mathematically formulate the history functions. First, it is assumed that the population of tumor cells in interphase is three times greater than in mitosis. Second, the slope of the line representing the growth of $T_M(t)$ cells is steeper due to the shorter duration of mitosis in the cell cycle. These assumptions are reflected in Figure \ref{historyfunction}.
The population of tumor cells in interphase and mitosis at the moment of detection is approximately three-fourths ($0.21\times10^{11}$) and one-fourth ($0.07\times10^{11}$) of the total population size, respectively. Additionally, the intrinsic delay, assumed to be $\tau=2$ days, is reflected in the history functions.

Following the outlined process and integrating the normalized initial distribution of cells over the tumor site, the initial population of different cell types is calculated as:
$$N(x,0)=0.2 \rm exp(-2x^2), \int _{-0.5}^{0.5} 0.2  \rm exp(-2x^2)\;\mathrm{d}x= 0.171125\approx0.17$$
$$I(x,0)=0.375-0.235 \rm Sech^2(x), \int _{-0.5}^{0.5} 0.375-0.235 \rm Sech^2(x)\;\mathrm{d}x= 0.157805\approx0.15$$
As a result, the initial population of different types of cells is:\\
$\left (N(x,0),~ T_I(x,0),~ T_M(x,0),~ I(x,0)\right)=(0.171125,~ 0.209571,~ 0.069857,~ 0.157805)$.
These history functions will now be used to analyze the dynamics of the tumor-immune interaction.

\section{Effectiveness of the immune system}\label{Effectiveness of the immune system}
The immune system plays a crucial role in cancer formation, progression, and treatment in tumor-bearing hosts. Under certain conditions, as explained in Section \ref{immunesurveillance}, tumor cells evade immune surveillance, continuing to proliferate and leading to the spontaneous outbreak of tumor growth. This immune dysfunction can be attributed to tumor cell resistance to cytokines, disruptions in immunoregulation induced by tumor cells, or antigen masking on tumor cell surfaces.
As shown in \cite{ansarizadeh2017modelling}, the strength of the immune system critically affects the interactions between host tissue cells and tumor cells. A healthier host exhibits greater potential for immune cells to eliminate tumor cells, resulting in better responsiveness to treatment.

According to \cite{de2003dynamics}, $\dfrac{s}{d}$ can be considered as an indicator of immune system strength.
In the following \textit{in silico} experiments, the value of $d$ is kept unchanged and thus different values of parameter $s$ illustrate different levels of immune strength of the tumor-bearing host.
The mechanisms of the immune system remain incompletely understood, with some claims suggesting paradoxical responses to tumor presence \cite{de2006paradoxical}. Thus, mathematical modeling offers valuable theoretical and preclinical insights, paving the way for clinical breakthroughs once hypotheses are validated experimentally. To explore this, some \textit{in silico} experiments are conducted to examine the role and significance of the parameter $s$ in tumor-immune interactions.

\begin{figure}[htbp]
\centering
\includegraphics[width=0.5\textwidth]{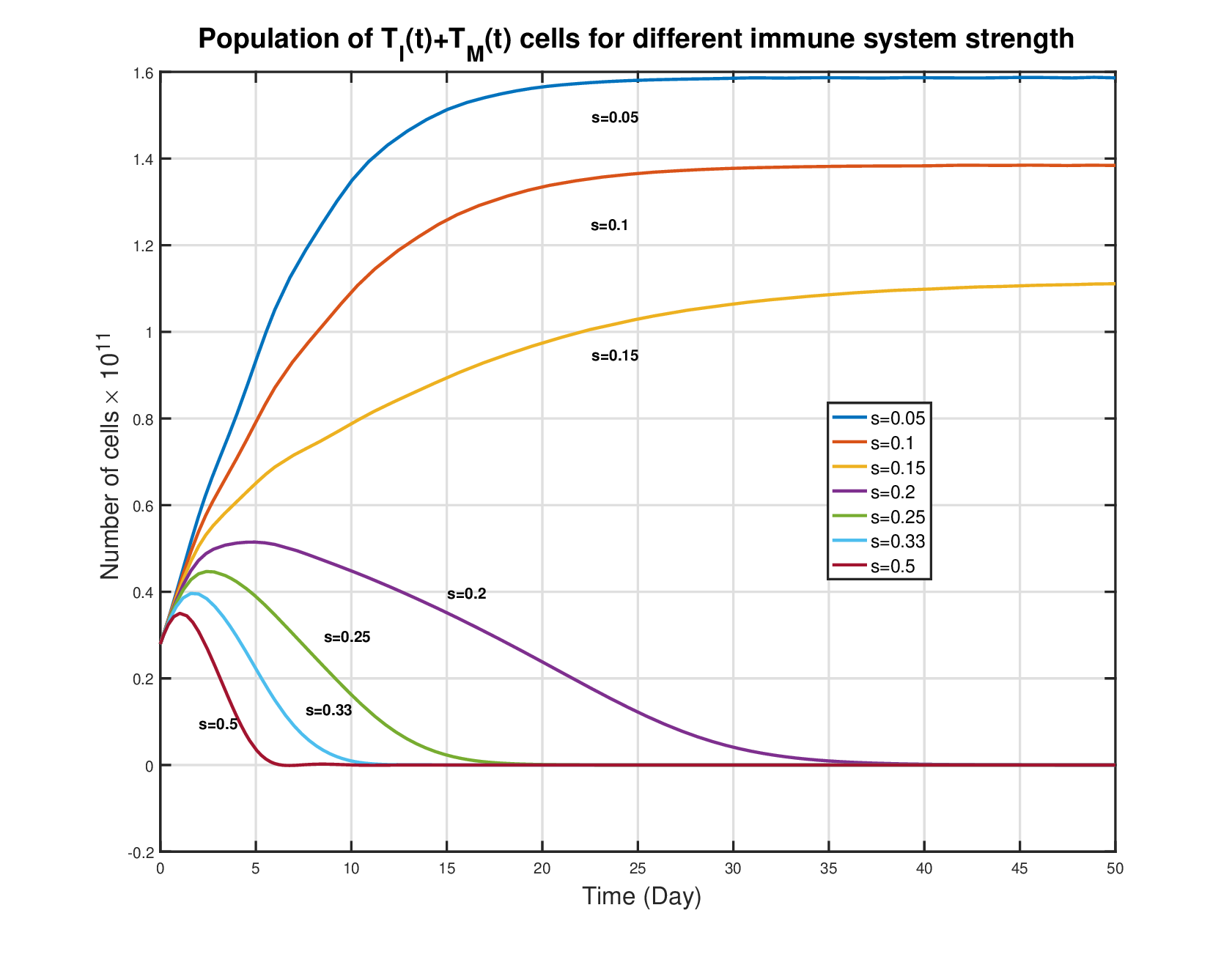}
\caption{The effect of the constant influx of immune system cells, $s$, on the regression or progression of tumor cells as a natural response of the host, under varying levels of immune system strength, without the administration of chemotherapeutic agents.}
\label{Nodrug}
\end{figure}

Figure \ref{Nodrug} demonstrates that for lower values of the parameter $s$, representing the constant influx of immune system cells in the absence of tumor cells, the tumor cell population increases. In contrast, for higher values of $s$, the host's immune system can suppress and even eradicate the tumor cells. This suggests that techniques aimed at enhancing immune system strength, such as immunotherapy and vaccine therapy, can play a crucial role in combating tumor cells. As detailed in previous sections, with $s=0.33$, the system remains stable, indicating that immune surveillance is sufficiently robust to recognize, fight, and eliminate tumor cells.

\subsection{Cells evolution in different levels of the immune system strength}\label{immunesurveillance}
In this section, the evolution of cells under various conditions is analyzed. By keeping all parameters in equations \eqref{15}-\eqref{18} consistent with the values in Table \ref{parameterset}, except for parameter $s$, the results highlight the influence of immune system strength on the dynamics of all cell types.

\begin{figure}[htbp]
\centering
\subfloat[s=0.33] {\label{ss} \includegraphics[width=0.45\textwidth]{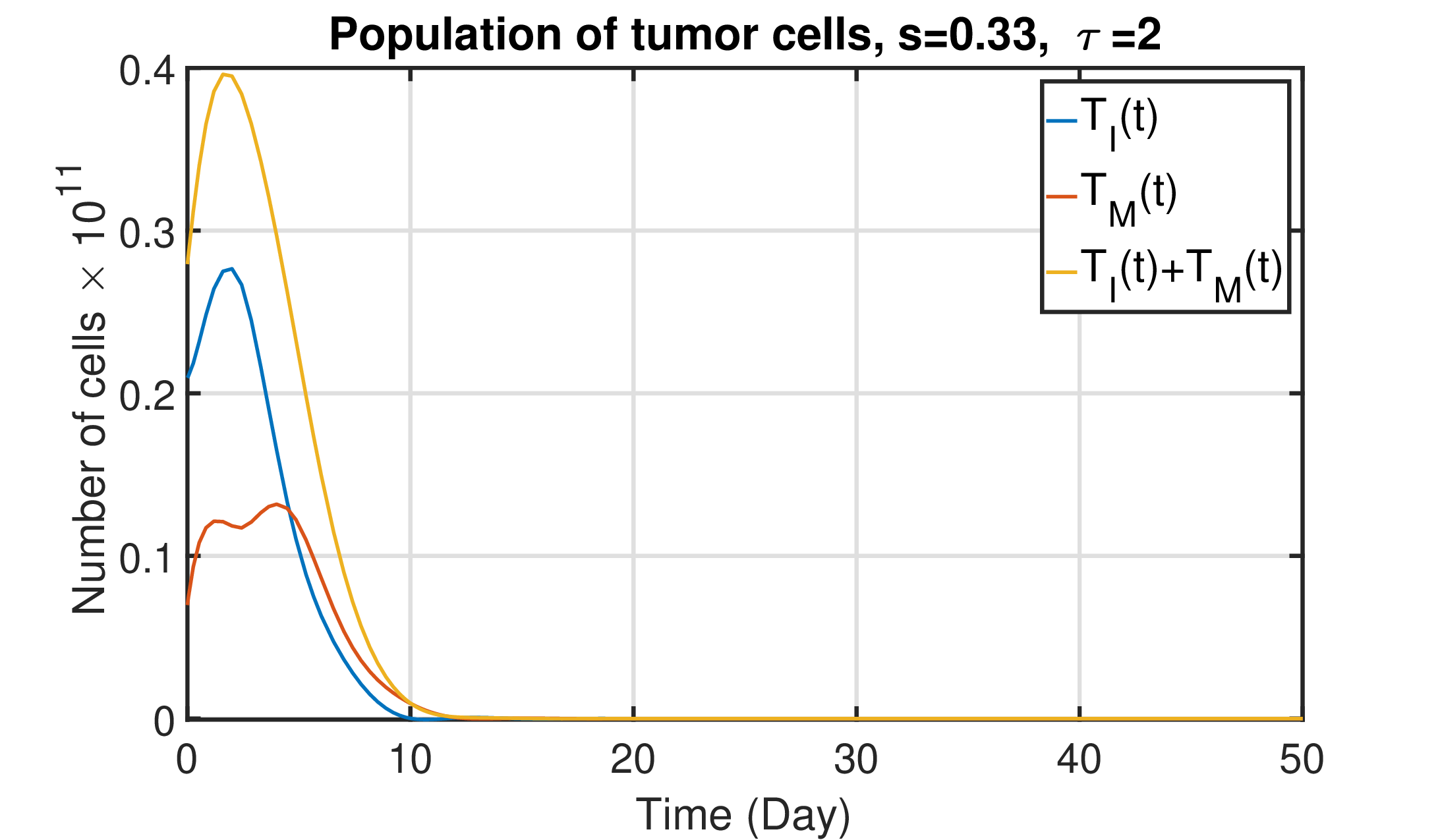}}
\subfloat[s=0.1] {\label{s} \includegraphics[width=0.45\textwidth]{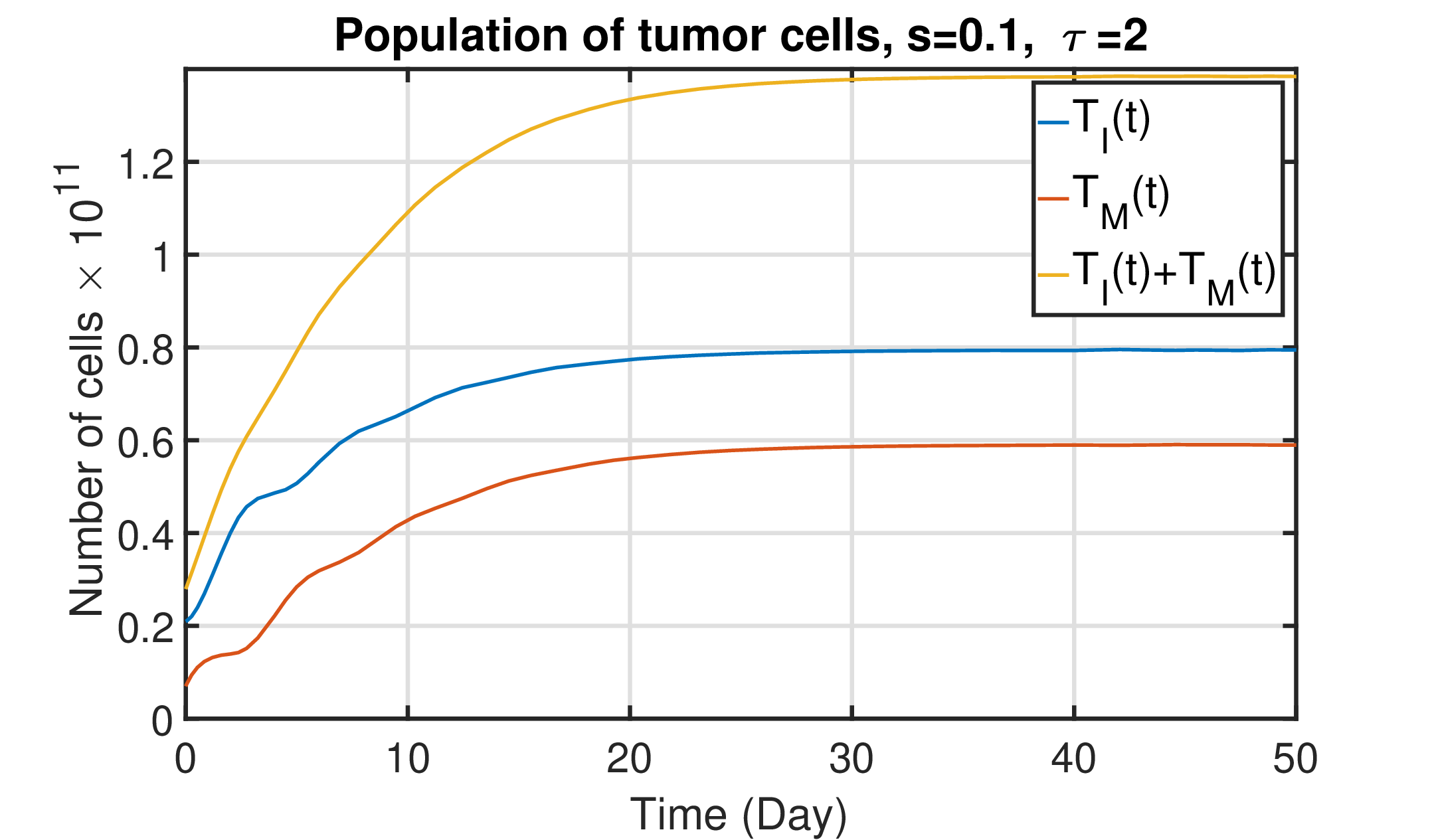}}
\caption{The impact of the immune system on the regression or progression of tumor cells under the given parameter set, with variations in parameter $s$.}
\label{TITMNodrug}
\end{figure}

\begin{figure}[htbp]
\centering
\subfloat[s=0.33]{\label{evloutions033} \includegraphics[width=0.45\textwidth]{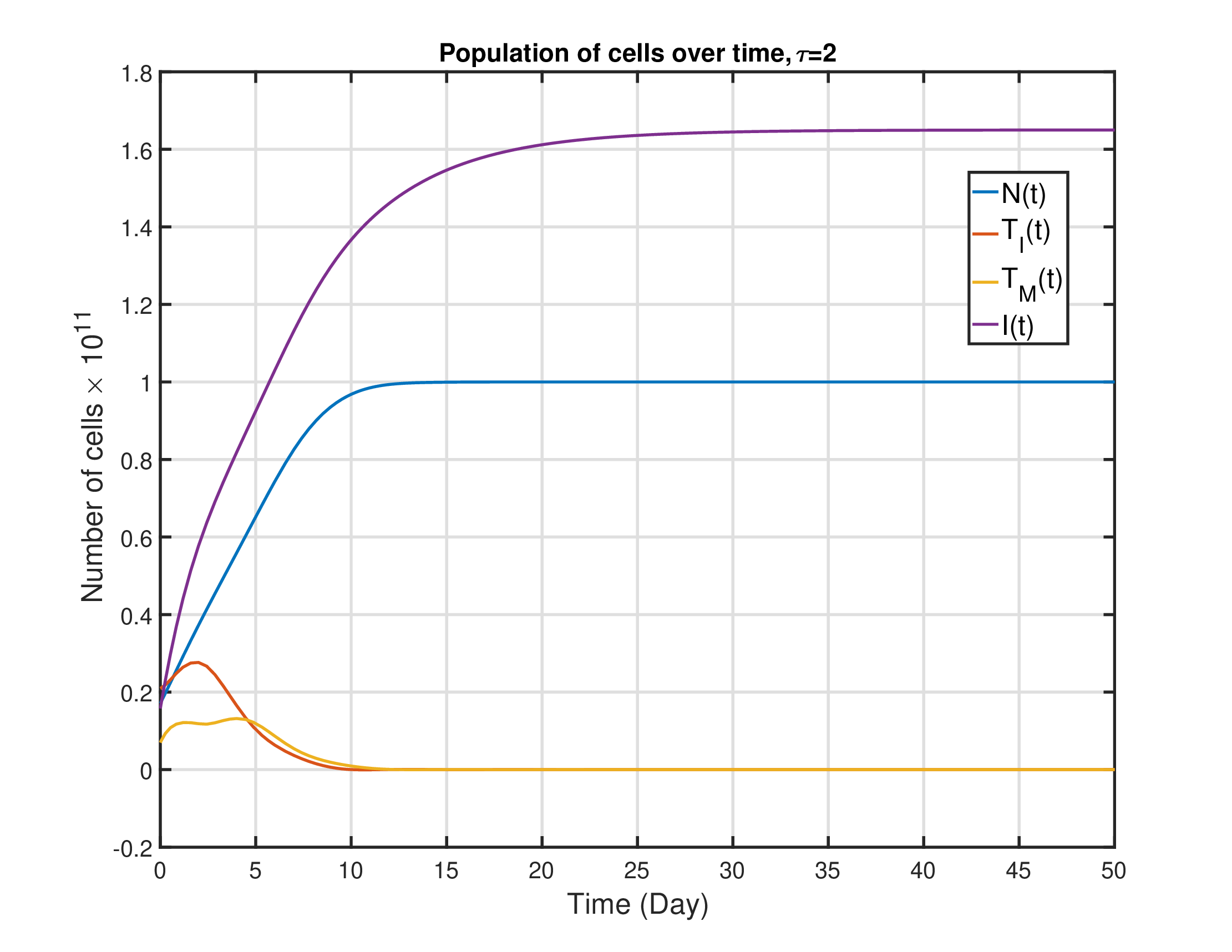}}
\subfloat[s=0.1]{\label{evloutions=01} \includegraphics[width=0.45\textwidth]{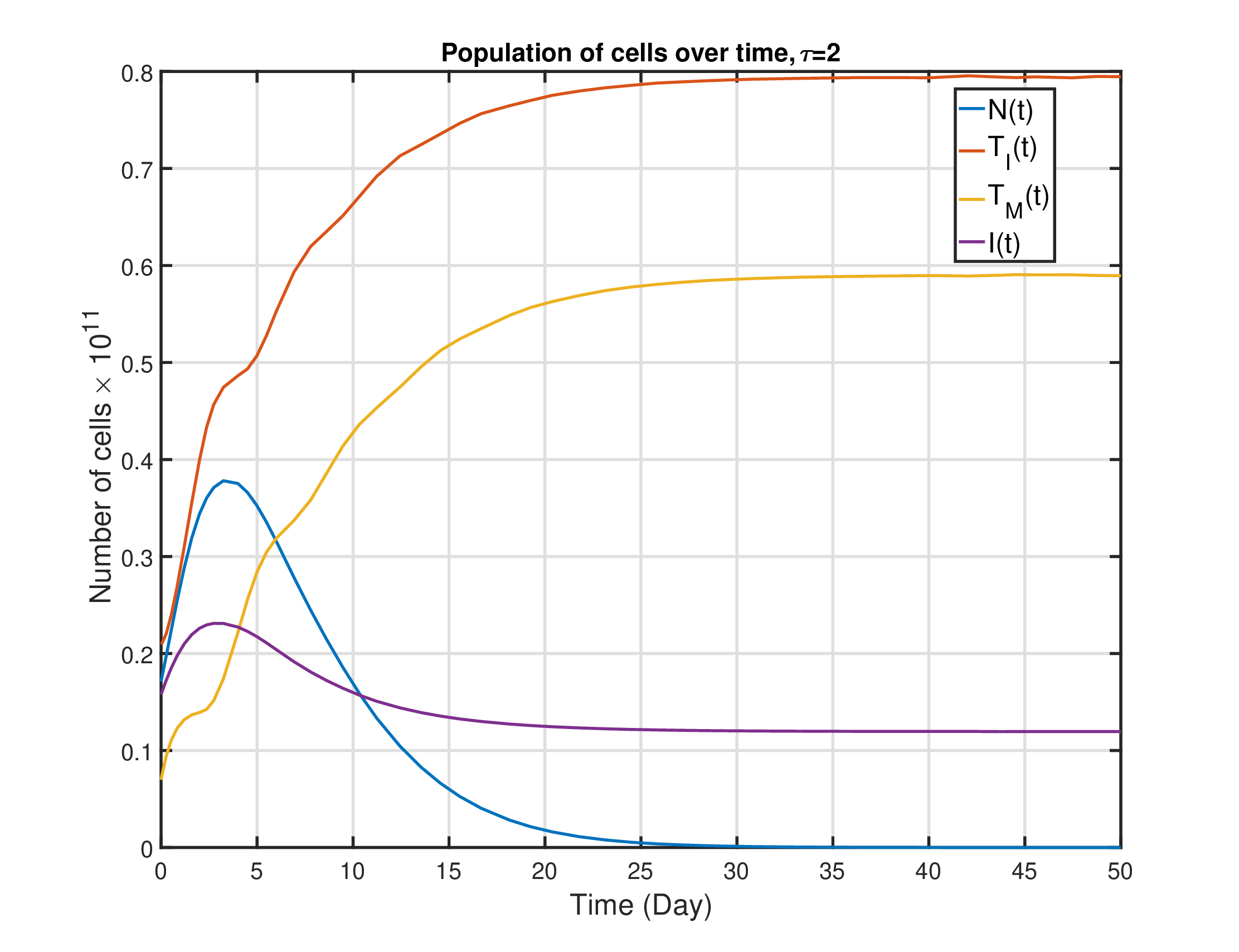}}
\caption {Evolution of normal, tumor, and immune cells in tumor-bearing hosts with healthy ($s=0.33$) and compromised ($s=0.1$) immune systems.}
\label{xx}
\end{figure}

Figures \ref{TITMNodrug} and \ref{xx} present the mathematical results for healthy ($s=0.33$) and compromised ($s=0.1$) immune systems, highlighting tumor cell dynamics and the dynamics of all cell types, respectively. Figure \ref{ss} demonstrates that a strong immune system can eradicate tumor cells, whereas Figure \ref{s} shows that a compromised immune system allows tumor cells to dominate the tumor site. While Figure \ref{xx} illustrates the dynamics of all cell types, Figure \ref{TITMNodrug} provides a more detailed view of tumor cell proliferation over time, including their different phases and total population.

These mathematical experiments indicate that with robust immune surveillance, treatment may not be required, as the host can independently eliminate tumor cell mutations. However, in reality, this ideal scenario is unlikely and may instead lead to tumor cell dormancy, delaying the tumor's manifestation.

\subsubsection{Routh–Hurwitz criteria for compromised immune system}

For the vulnerable immune system case with parameter $s=0.1$, equilibrium points of the system of equations \eqref{15}-\eqref{18} were determined. The tumor-free scenario yields the equilibrium point $P^o(1,~0,~0,~0.5)$. In the coexistence scenario, among the mathematically possible equilibrium points, only $P^*(6.546\times10^{-7},~0.7944,~0.5900,~0.1195)$ satisfies the biological and mathematical conditions outlined in Section \ref{Points of equilibria in coexisting case}. The Routh–Hurwitz conditions for this equilibrium point under the compromised immune system are provided in Table \ref{RHcompromised}.

\begin{table}[htp]
\caption{Routh-Hurwits condition for equilibrium point of in a compromised immune system.}
\centering
\begin{tabular} {c|c c c c}\Xhline{1.2pt}
\diagbox[width=10em]{\thead {Equilibrium \\points}}{\thead {RH \\criteria}}&\thead {$P_1>0$}&\thead {$P_3>0$}&\thead {$P_4>0$}&\thead {$P_1P_2P_3>P^{2}_3+P^{2}_1P_4$}\\ \Xhline{1.2pt}
$P^o$&$3.47$&$-0.358$&$-0.1353$&$-2.792>-1.5019$\\ \Xhline{1.2pt}
\end{tabular}
\label{RHcompromised}
\end{table}
\begin{table}[htp]
\caption{Eigenvalues of the points of equilibria.}
\centering
\begin{tabular} {c|c c c c}\Xhline{1.2pt}
\diagbox[width=10em]{\thead {Equilibrium \\points}}{\thead {Eigenvalues}}&\thead {$\lambda_1$}&\thead {$\lambda_2$}&\thead {$\lambda_3$}&\thead {$\lambda_4$}\\ \Xhline{1.2pt}
$P^o$&$-2.53675 $&$ -1$&$-0.2$&$0.266749 $\\ \Xhline{1.2pt}
\end{tabular}
\label{eigenvaluest}
\end{table}
As evident, three of the four Routh–Hurwitz conditions are violated. Similarly, Table \ref{eigenvaluest} indicates a positive eigenvalue, further confirming the rejection of stability conditions. Thus, the tumor-free state is unstable, implying that, in the absence of strong immune surveillance, the initiated mutation shifts the system from a tumor-free state to a coexistence scenario. Consequently, tumor cells continue to proliferate, resulting in the formation of an undesirable mass, the tumor. Such situations necessitate the application of treatment.

\section{Therapeutic strategies}\label{Treatments strategies}
From now on, the focus is on a compromised or weak immune system, where tumor cell mutations persist and progress to form a tumor.
This study specifically selects Paclitaxel (generic name) or Taxol (brand name) as the "antineoplastic" or "cytotoxic" chemotherapy drug. Paclitaxel is administered intravenously, with infusion durations ranging from 30 minutes to 8 hours per chemotherapy session \cite{athawale2015chemotherapy}. Since this drug is used for primary breast cancer before metastasis, it is suitable for treating the early-stage tumor considered in this study.
The primary objective of the treatment is to minimize the tumor cell population. The dosage and frequency of chemotherapy administration critically influence the therapy's outcome. Drug scheduling, a significant area of research \cite{de2001mathematical}, is explored here using two common chemotherapy administration methods. Additionally, the results of \textit{in silico} experiments are verified to align with \textit{in vivo} observations.
For decades, specialists believed that the MTD approach was the most effective cancer treatment due to its high chemotherapy dosages. However, these high doses lead to significant short-term and long-term side effects that adversely impact the patient's quality of life. In response to these drawbacks, a novel method called metronomic chemotherapy has emerged, offering less severe side effects and more desirable outcomes.
The metronomic procedure, first used for treating human breast cancer in 2002, has proven to be significantly less toxic compared to MTD \cite{colleoni2002low}. In this study, \textit{in silico} experiments will be conducted to verify the proposed model's ability to simulate outcomes consistent with clinical observations. A key concept from the literature is that a single chemotherapy agent is insufficient to eradicate cancer. Finding an optimal dosage that balances minimal toxicity with effectiveness remains challenging \cite{hanahan2000less}. Researchers are exploring various regimens through clinical trials and mathematical modeling to identify the most effective metronomic approach.
In Section \ref{Aca}, a mathematical experiment is conducted to determine an appropriate and pragmatic drug dosage, yielding reasonable outcomes for the treatment.

\subsection{Administrating chemotherapeutic agents}\label{Aca}
\begin{figure}[htpb]
\centering
\subfloat[ Maximum tolerated dosage (MTD) approach.]{\label{mtd}\includegraphics[width=0.5\textwidth]{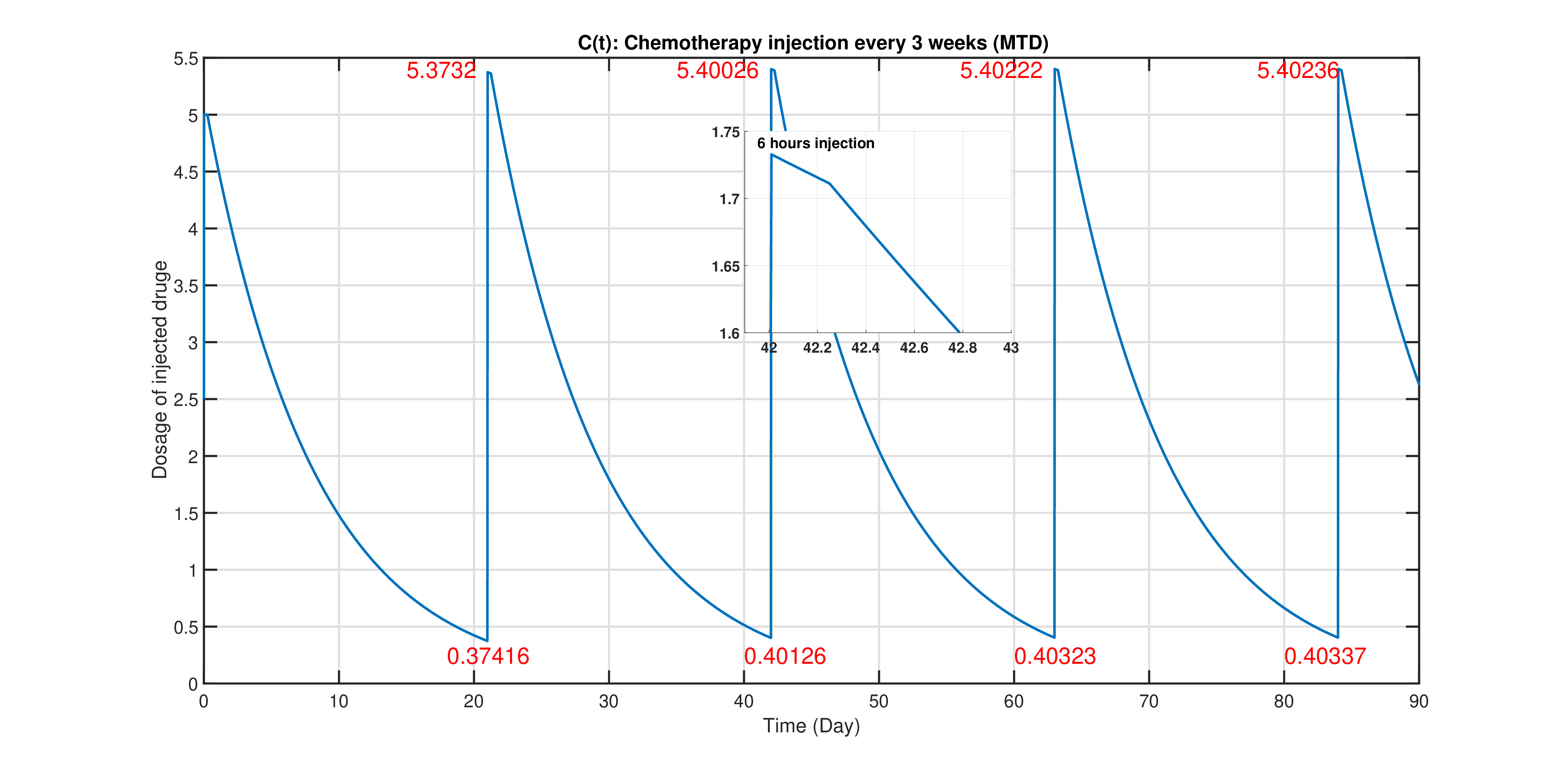}}
\subfloat[ Metronomic approach.]{\label{metro}\includegraphics[width=0.5\textwidth]{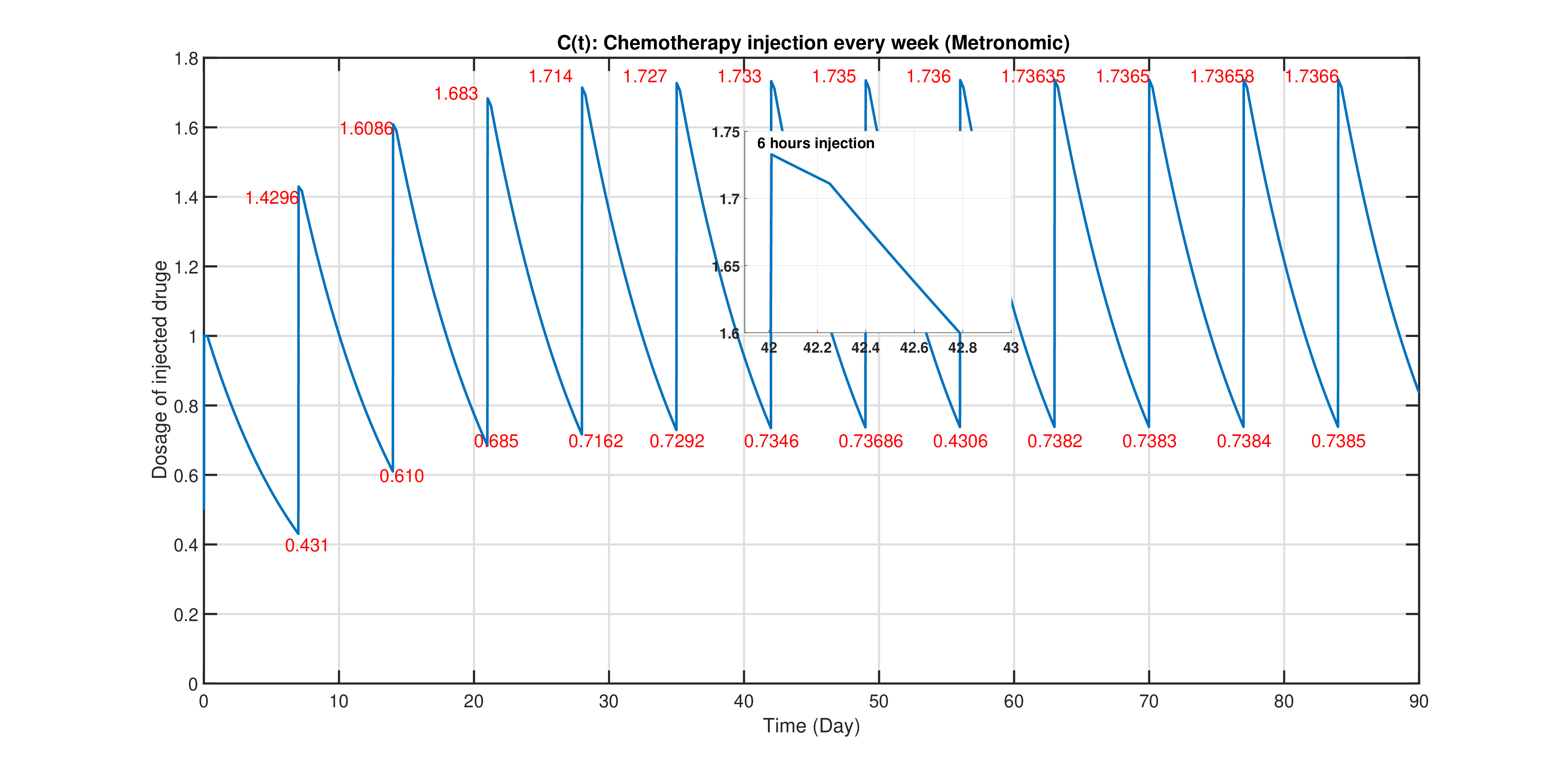}}~~~~~
\caption{Dosage and frequency of chemotherapeutic agent injections in MTD and Metronomic protocols.}
\label{chemotherapy}
\end{figure}

Figure \ref{mtd} depicts the typical dosage and frequency for chemotherapeutic agents over 3 months, with each chemotherapy infusion session lasting 6 hours and a 3-week recovery period, which is a common schedule. In our mathematical model, the MTD approach uses the same dosage and frequency, while the metronomic approach reduces the dosage to one-fifth and increases the infusion frequency to weekly, as shown in Figure \ref{metro}.
Figure \ref{chemotherapy} shows that the maximum dosage of the chemotherapeutic agent follows a monotonic increasing trend in both treatment protocols, as the leftover from the previous injection is added to the new dosage. The key observation is the cumulative percentage of the current dose in both chemotherapy approaches over the same treatment period: $8.05\%$ for MTD and $73.66 \%$ for metronomic. This suggests that more frequent administration of chemotherapeutic agents may result in a more effective response.

\subsection{Mathematical interpretation of the chemotherapeutic agents}
Mathematically, incorporating the effect of chemotherapeutic agents on tumor cell demise, in line with clinical observations, is crucial. Interestingly, contrary to expectations, increasing the dosage of chemotherapeutic agents does not necessarily lead to greater tumor cell eradication. Clinical studies indicate that at low concentrations, the effect is approximately linear, but beyond a certain threshold, increasing the dosage yields diminishing returns, leading to saturation. Considering this natural behaviour of chemotherapeutic agents, the following term can be used to model this pattern, where $i=N, ~T_I, ~T_M, ~I$, and $e$ is the exponential function:
\begin{align}
F(C(t))=a_i (1-e^{-C(t)}).
\label{FC}
\end{align}
The function $C(t)$, representing the dosage of chemotherapeutic agents over time, is plotted in Figure \ref{chemotherapy}. 
Using the Taylor series expansion, it is evident that for small $C(t)$, $F(C(t))$ behaves linearly, while as $C(t)$ increases, the upper limit of $F(C(t))$ approaches a constant value. 

Chemotherapeutic agents are unable to specifically target tumor cells, leading to the unintended destruction of both immune system cells and healthy cells in the tumor-bearing host \cite{zitvogel2008immunological}. However, due to their antineoplastic properties, chemotherapeutic agents exhibit a higher rate of tumor cell eradication compared to other cell types.
In the mathematical model of chemotherapy, the coefficient $a_i$ in equation \eqref{FC} accounts for the varying rates of cell demise across different cell types. The selection of $a_i$, representing the fraction of cell kill, requires imposing biological constraints derived from empirical research to ensure realistic and accurate modeling.
According to \cite{de2003dynamics}, the possible range for these coefficients is as $0\leq a_i \leq0.5,$ and $a_N\leq a_I\leq a_T$.

An ideal situation with $(a_T\approx1,~ a_I\approx0, ~a_N\approx0)$ would lead to unrealistic outcomes, but we aim to model a real scenario. Tumor cells are categorized into interphase and mitosis phases, with different fraction cell kill coefficients assigned to each. Medical observations suggest that chemotherapeutic agents, while not perfectly targeted, tend to kill more malignant cells at a higher rate. Therefore, a reasonable assumption is $a_{T_M} \leq a_{T_I}$ and $a_N \leq a_I \leq a_{T_M} \leq a_{T_I}$. 
Hence, the fraction of cell kill for different cell types can be chosen as $a_N=0.1$, $a_{T_I}=0.3$, $a_{T_M}=0.25$, and $a_I=0.2$.

\subsection{Influence of the chemotherapy on the evolution of cells}
As discussed in Section \ref{Treatments strategies}, a combination of therapeutic approaches is required to effectively eliminate tumor cells. Therefore, it is unrealistic to expect chemotherapy alone to completely eradicate the tumor. While significant regression in the tumor cell population is observed with chemotherapy, the tumor is not fully annihilated.
Introducing the effect of chemotherapy through equation \eqref{FC} modifies the system of equations \eqref{15}-\eqref{18} as follows:
\begin{align}
\label{}
\dot  N(t)&=  N(t) \left(1-\frac{ N(t)}{b_1}\right)- c_1T_I(t) N(t)- c_2 T_M(t) N(t) -a_N (1-e^{-C(t)})N(t),\\
\end{align}
\begin{align}
\dot T_I(t)&=  r T_I(t) \left(1-\frac{ T_I(t)}{b_2}\right)+2 \alpha_IT_M(t)-\alpha_MT_I(t-\tau)-\beta_IT_I(t)-c_3  I(t)T_I(t)-c_4 N(t) T_I(t)\nonumber \\
&-a_{T_I} (1-e^{-C(t)})T_I(t),
\label{}
\end{align}
\begin{align}
\label{}
\dot T_M(t)&= \alpha_MT_I(t-\tau)-\alpha_IT_M(t)-\beta_MT_M(t)-c_5 I(t) T_M(t)-c_6 N(t) T_M(t)-a_{T_M} (1-e^{-C(t)})T_M(t),
\end{align}
\begin{align}
\label{}
\dot I(t)&= s+\rho\frac{ I (t) \left( T_I(t)+T_M(t)\right)}{\alpha+T_I(t)+T_M(t)}-c_7 T_I(t)I(t)-c_8T_M(t)I(t)-d I(t)-a_I (1-e^{-C(t)})I(t).
\end{align}

\subsubsection{Outcomes of MTD and metronomic approaches}

\begin{figure}[htbp]
\centering
\subfloat[ Taking MTD protocol]{\label{MTD}\includegraphics[width=0.45\textwidth]{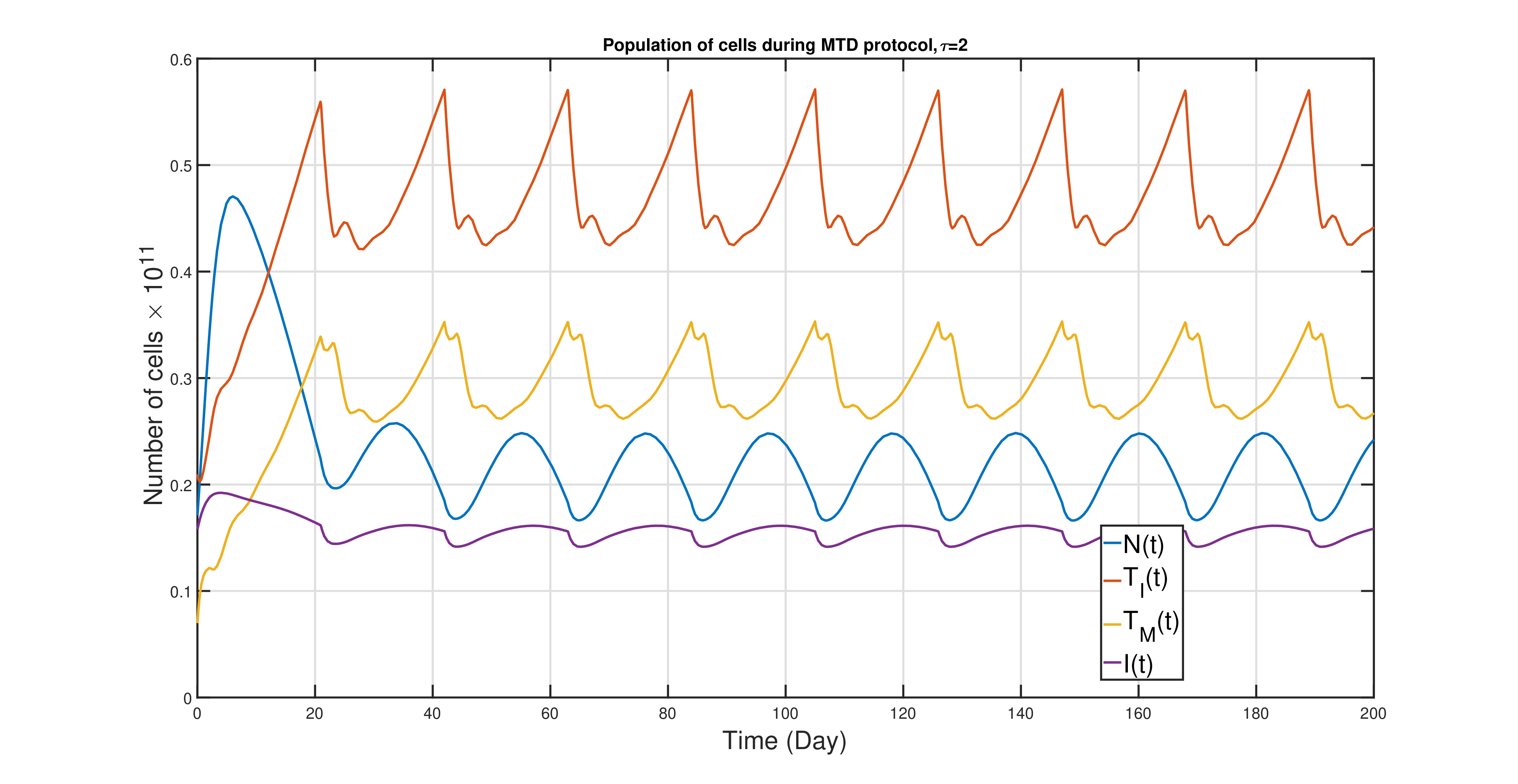}}
\subfloat[Taking metronomic protocol ]{\label{Metronomic}\includegraphics[width=0.45\textwidth]{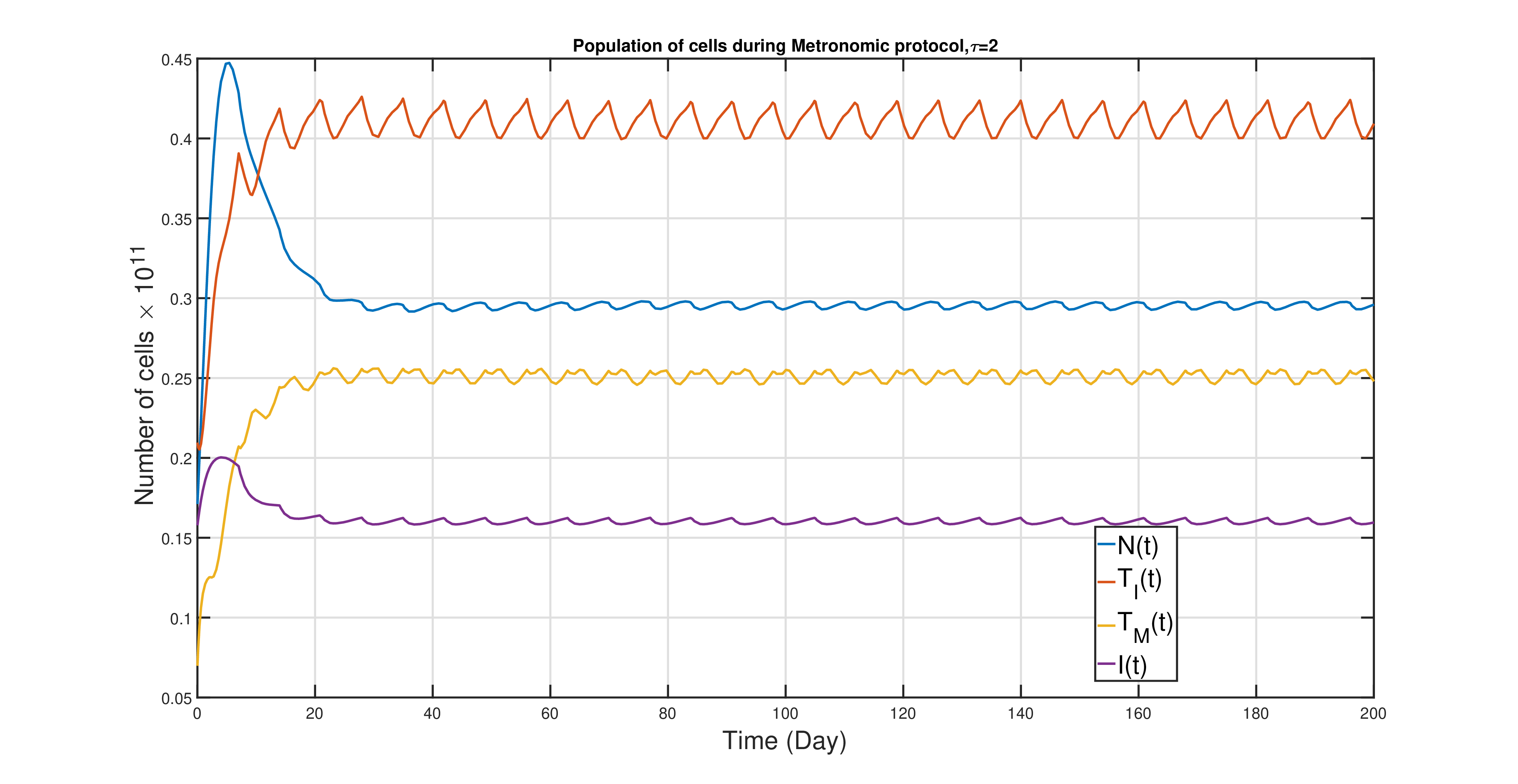}}
\caption{Outcomes of chemotherapy using MTD and metronomic approaches over a 6-month treatment period.}
\label{twotreatment}
\end{figure}

Comparing Figure \ref{evloutions=01}, which shows the evolution of different cell types post-mutation in the tumor-bearing host, with Figures \ref{twotreatment}, highlights the efficacy of MTD and metronomic chemotherapy regimens and contrasts their desirable effects.

In the MTD procedure, over more than 6 months of treatment, the maximum population of tumor cells in interphase, $T_I$, in a compromised tumor-bearing host, as shown in Figure \ref{evloutions=01}, decreases from $0.79 \times 10^{11}$ to $0.57 \times 10^{11}$, while the population of tumor cells in mitosis, $T_M$, regresses from $0.59 \times 10^{11}$ to $0.35 \times 10^{11}$, as illustrated in Figure \ref{MTD}.
During the same period, the metronomic chemotherapy technique leads to a greater reduction in tumor cell populations in both interphase and mitosis compared to the MTD regimen. Specifically, the maximum populations of $T_I$ and $T_M$ cells decrease to $0.42 \times 10^{11}$ and $0.25 \times 10^{11}$, respectively.

Theoretically, treatment is only necessary when the immune system is compromised ($s=0.1$), as the host's immune response is insufficient to eradicate the tumor. Therefore, Figure \ref{evloutions=01} illustrates the evolution of cells without chemotherapy, providing a baseline for comparison with the cell populations post-chemotherapy, shown in Figure \ref{twotreatment}. For example, the population of normal cells decreases from $3.78 \times 10^{10}$ to approximately zero. This demonstrates that, without chemotherapy and with a weakened immune system, tumor cells dominate and healthy cells are lysed or destroyed.
With chemotherapy administration, the population of normal cells fluctuates between $1.6 \times 10^{10}$ and $4.7 \times 10^{10}$ under the MTD technique. In contrast, the metronomic approach results in a variation in the normal cells' population from $4.47 \times 10^{10}$ to $1.7 \times 10^{10}$, which is more favourable for tissue preservation compared to the MTD method.
Figure \ref{twotreatment} clearly shows that while chemotherapy reduces the tumor cell population, it cannot completely eradicate the cells and merely controls their proliferation. Hence, incorporating adjuvant therapies is essential to achieve complete cancer cell destruction.

To better understand the beneficial effects of the administered chemotherapy techniques, Table \ref{percentage} presents the percentage regression in the tumor cell population for both chemotherapy protocols. These values are based on the average tumor cell population. Jeff's phenomenon is observed at the beginning of both chemotherapy regimens, reflecting an initial unexpected response to therapy due to the presence of drug-resistant subpopulations of tumor cells.
As demonstrated in \cite{ansarizadeh2017modelling}, these subpopulations exhibit different and even contradictory responses in the initial stages of treatment. Thus, it may be prudent to disregard the initial decline in tumor cells immediately following treatment initiation.
\begin{table}[htpb]
\caption{Percentage of regression among tumor cells after chemotherapy}
\centering
\begin{tabular} { c c c }\\ \Xhline {1.2pt}
\multicolumn{1}{c|}{\multirow{3}{*}{\diagbox[width=10em]{\thead{Treatment \\ protocol}}{\thead{Population \\ of cells}}}}&
\multicolumn{2}{c}{\multirow{3}{*}\makecell{{\thead{\\ Regression in tumor cells\\ during interphase and mitosis}}}}\\
&\multicolumn{1}{|c}{}&\\
&\multicolumn{1}{|c}{{\thead{$T_I$}}}&{\thead{$T_M$}}\\ \Xhline{1.2pt}
\makecell {MTD\\(\small{Figure \ref{MTD}})}&\multicolumn{1}{|c}{$30.8\%$}&\multicolumn{1}{|c}{$47.5\%$}\\ \hline
\makecell {Metronomic\\(\small Figure \ref{Metronomic})}&\multicolumn{1}{|c}{$48.5\%$}&\multicolumn{1}{|c}{$57.7\%$}\\ \Xhline{1.2pt}
\end{tabular}
\label{percentage}
\end{table}
Comparing the first and second columns of Table \ref{percentage}, it is evident that $T_I$ cells are less responsive to treatment, despite $a_I$, the fraction cell kill for tumor cells in interphase, being the highest. This can be explained by the higher malignancy of tumor cells in interphase compared to mitosis. Furthermore, the metronomic approach shows a distinct advantage over the MTD method, as it maintains a higher ratio of healthy cells to $T_M$ cells. This suggests that more frequent chemotherapy injections with lower dosages are more beneficial in preserving normal cells.

\section{Importance of immune system strength on the efficiency of treatment}
\begin{figure}[ht]
\centering
\subfloat[ ]{\label{mtdn5}\includegraphics[width=.35\textwidth]{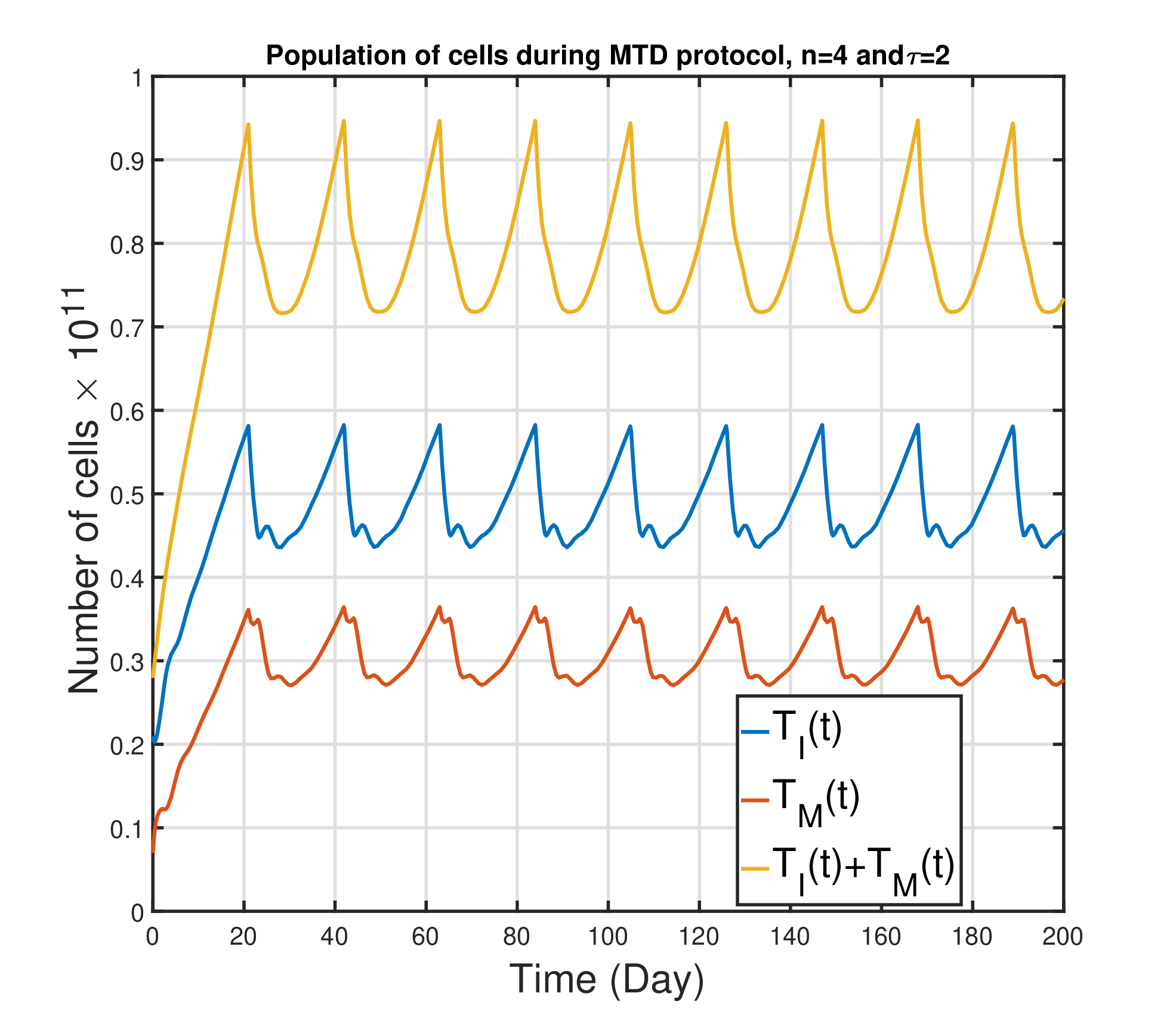}}~~~~~
\subfloat[ ]{\label{mtdn05}\includegraphics[width=.35\textwidth]{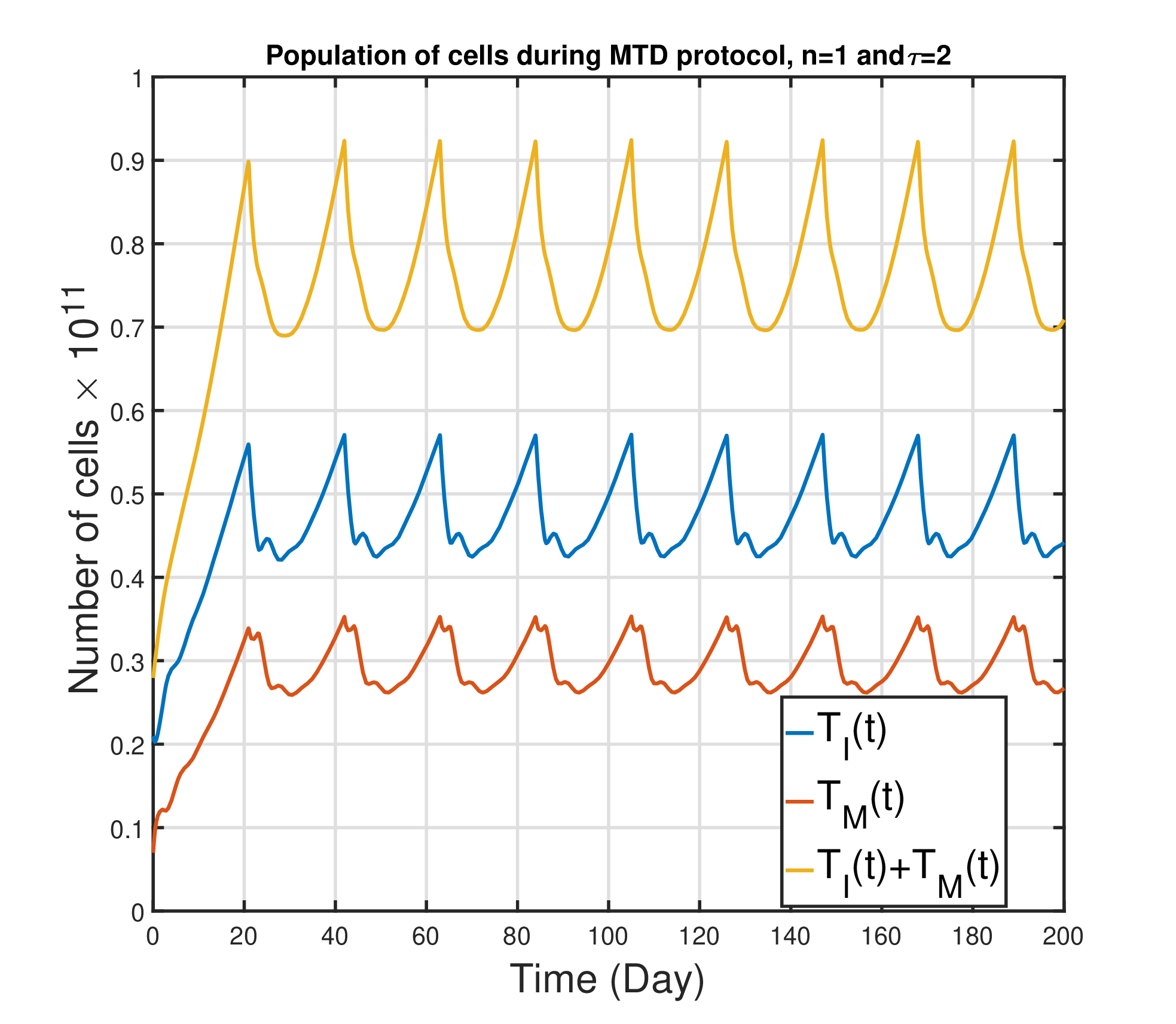}}\\
\subfloat[ ]{\label{n=5}\includegraphics[width=.35\textwidth]{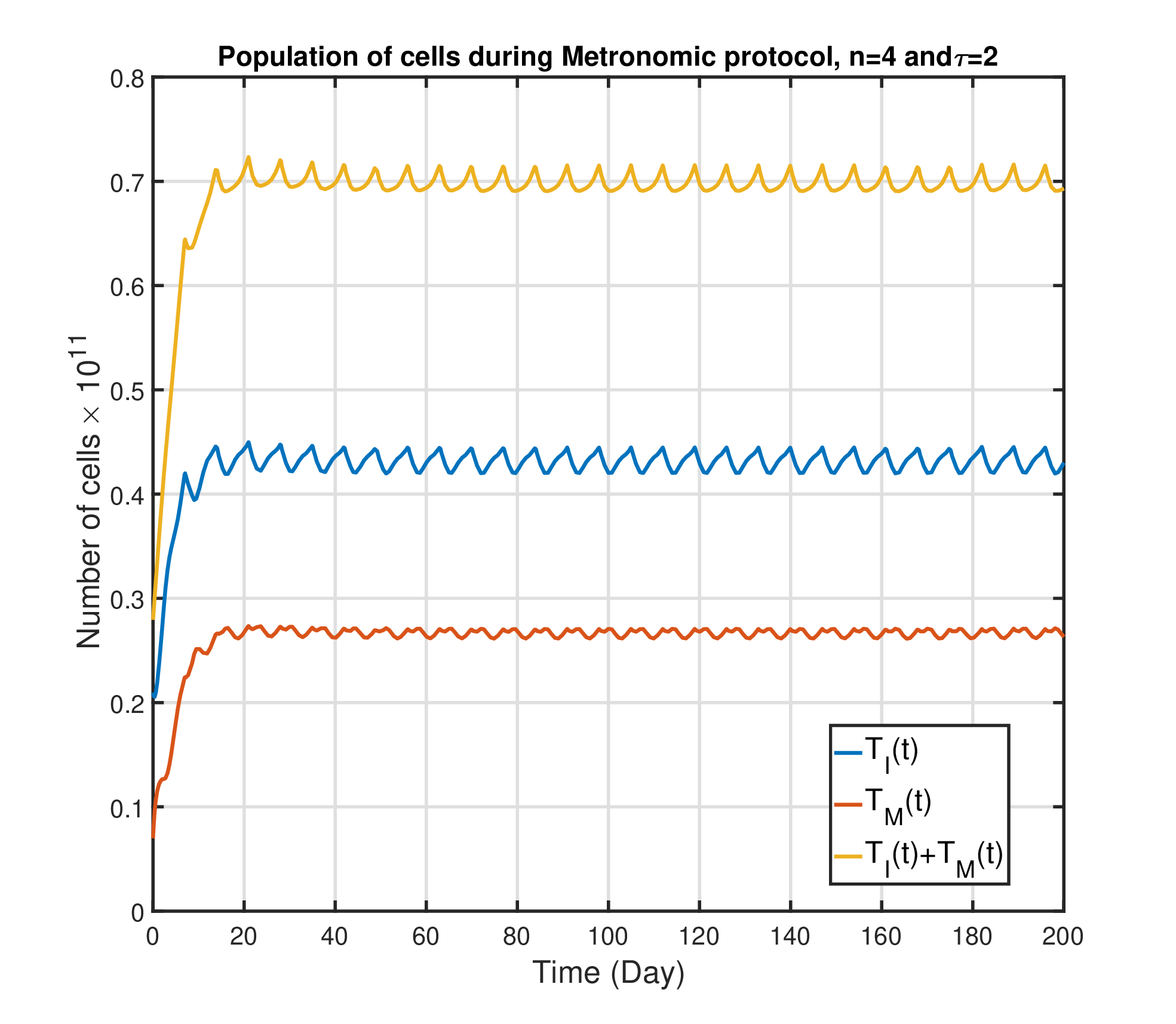}}~~~~~
\subfloat[ ]{\label{n=05}\includegraphics[width=.35\textwidth]{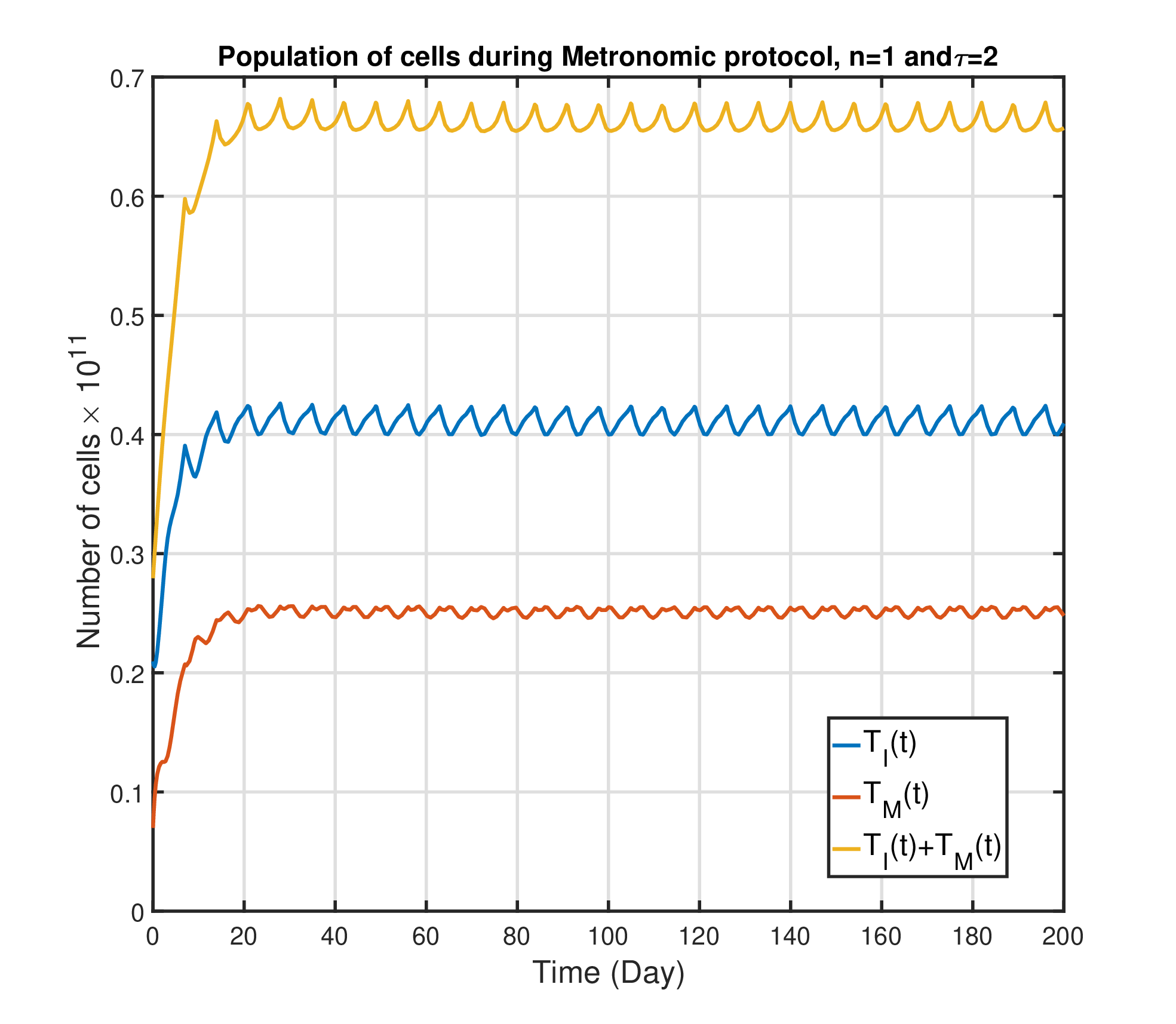}}
\caption{Regression of tumor cells following MTD and metronomic chemotherapy approaches for different values of parameter $n$, specifically $n=4$ and $n=1$.}
\label{Diffn}
\end{figure}

In this section, the effects of chemotherapy are examined for two extreme values of the parameter $n$, as described in equation \eqref{10} and Figure \ref{NS}, through a mathematical experiment. Figure \ref{Diffn} shows the regression in tumor cell population under both MTD and metronomic chemotherapy protocols. For the MTD method, comparing Figures \ref{mtdn5} and \ref{mtdn05} reveals that smaller values of $n$ result in greater regression of tumor cells, particularly when the total population of tumor cells ($T_I + T_M$) is considered. A similar pattern is observed in the metronomic method by comparing Figures \ref{n=5} and \ref{n=05}. These findings suggest that smaller values of $n$ lead to more effective chemotherapy outcomes, supporting the hypothesis proposed in Figure \ref{NS} that lower values of $n$ correlate with stronger immunogenicity. Additionally, higher values of $n$ indicate a slower decay in tumor cell population, suggesting that chemotherapy alone may only slightly reduce tumor cell regrowth after each treatment.
To gain a deeper understanding of chemotherapy dynamics, we examine the fluctuation in tumor cell population. This fluctuation appears to be an intrinsic characteristic of tumor cells, as they regrow after each chemotherapy session. While normal cells use the recovery period to heal from the side effects of chemotherapeutic agents, tumor cells begin proliferating again immediately.

\begin{figure}[htbp]
\centering
\subfloat{\label{bp}\includegraphics[width=0.43\textwidth]{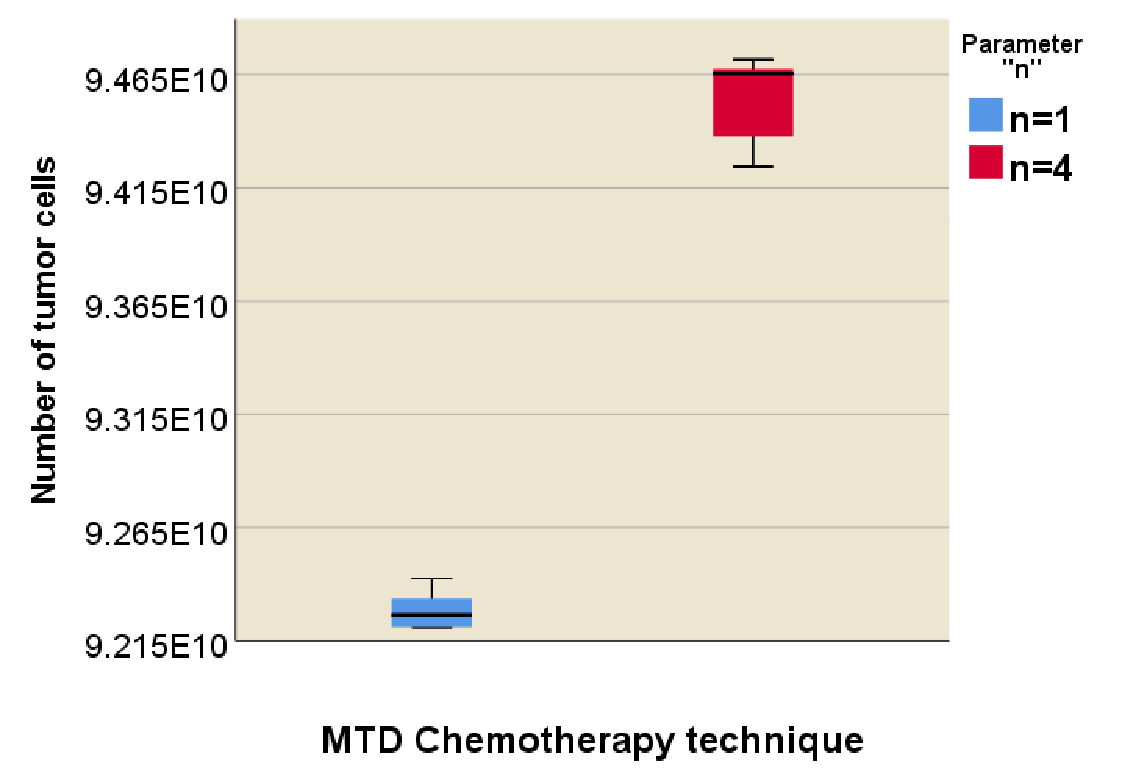}}
\subfloat{\label{bpm}\includegraphics[width=0.43\textwidth]{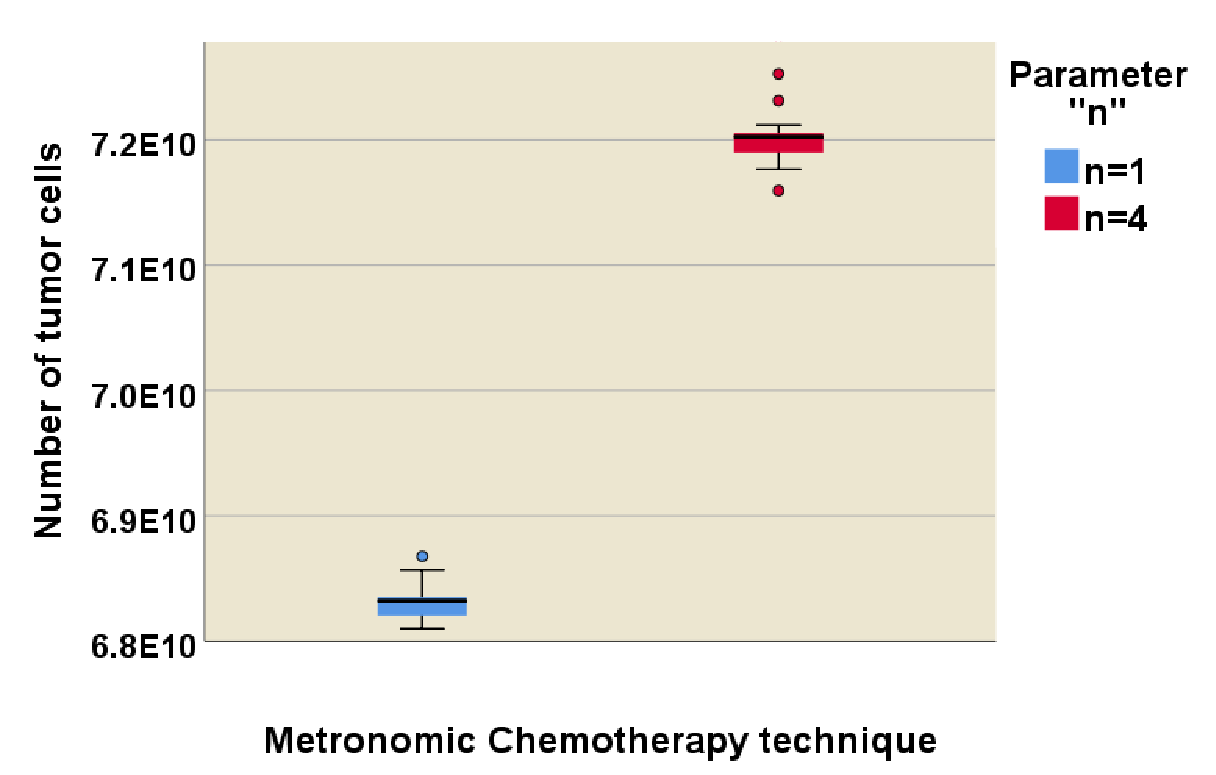}}
\caption{ Variations of the maximum points in the population of tumor cells as the outcomes of chemotherapy for MTD and metronomic approaches for healthy and compromised hosts, illustrated by $n=1$ and $n=4$, respectively.}
\label{boxplot}
\end{figure}

An innovative \textit{in silico} experiment determines the tumor cell population at each extremum point for the first time, revealing fluctuations that reflect overall patient condition consistency. As shown in Figure \ref{boxplot}, the metronomic technique results in fewer oscillations in tumor cell population compared to the MTD method, with a smaller interquartile range (IQR) in the box plots of Figure \ref{bpm} compared to those in Figure \ref{bp}. Biologically, more frequent treatment sessions with shorter intervals inhibit angiogenesis, despite lower dosages, supporting the anti-angiogenesis effect of the metronomic approach.
Furthermore, the model suggests that the metronomic procedure offers greater stability in patient health, with less variation in tumor cell population. Notably, the interquartile range (IQR) of the box plot is smaller with a stronger immune system, as indicated by smaller values of parameter $n$, further supporting the findings from Sections \ref{ex2} and \ref{ex1}. The outliers in the box plots represent the initial unexpected behaviour of the host, likely due to Jeff's phenomenon.

\section{Sensitivity analysis of the proposed model} \label{Sensitivity analysis of the proposed model}
This section conducts a sensitivity analysis to identify key parameters influencing the model. It evaluates how perturbations in quantitative factors (e.g., inputs, initial conditions, parameters) or variations in qualitative factors (e.g., structure, connectivity, or submodels) affect the model's behavior.
The local sensitivity of the proposed model, represented by the system of equations \eqref{15}-\eqref{18}, is analyzed in the presence of chemotherapeutic agents. In this analysis, parameters are varied individually around their nominal values, as provided in Table \ref{parameterset}. As shown in Figure \ref{xx}, $s=0.1$ represents a compromised immune system requiring chemotherapy. Therefore, the fixed point for this local sensitivity analysis is determined using the parameters in Table \ref{parameterset}, with $s$ set to $0.1$.

A widely used technique in sensitivity analysis is the sensitivity index ($SI$), where one parameter is varied over a specific range while others remain unchanged. The formula for $SI$ is given as $SI=\dfrac{D_{Max}-D_{min}}{D_{Max}}$, where $D_{Max}$ and $D_{min}$ represent the maximum and minimum outputs, respectively, when a parameter is altered individually. Tables \ref{SMTD} and \ref{SMetronomic} present the percentage variations in the populations of all interacting cell types, considering a $\pm 1\%$ deviation in parameter values one at a time.
\begin{table}[htp]
\caption{Sensitivity of the MTD approach to model parameters.}
\centering
\begin{tabular} {c|c c c c}\Xhline{1.2pt}
\diagbox[width=10em]{\thead {Parameters}}{\thead {Types of \\cells}}&\thead {$N$}&\thead {$T_I$}&\thead {$T_M$}&\thead {$I$}\\ \Xhline{1.2pt}
$b_1$&\small{$1.936$}&\small{$-0.0479$}&\small{$-0.0589$}&\small{$0.0312$}\\
$c_1$&\small{$-2.226$}&\small{$0.0457$}&\small{$0.0216$}&\small{$-0.0403$}\\
$c_2$&\small{$-1.243$}&\small{$0.037$}&\small{$0.0837$}&\small{$-0.0298$}\\
$r$&\small{$-3.623$}&\small{$1.39$}&\small{$1.77$}&\small{$-0.882$}\\
$b_2$&\small{$-3.775$}&\small{$2.172$}&\small{$2.488$}&\small{$-1.006$}\\
$\alpha_I$&\small{$-2.462$}&\small{$1.446$}&\small{$0.1724$}&\small{$-0.152$}\\
$\alpha_M$&\small{$-1.357$}&\small{$0.2069$}&\small{$2.272$}&\small{$-0.3725$}\\
$\beta_I$&\small{$0.7745$}&\small{$-0.3204$}&\small{$-0.3258$}&\small{$0.199$}\\
$c_3$&\small{$1.115$}&\small{$-0.540$}&\small{$-0.685$}&\small{$0.2887$}\\
$c_4$&\small{$0.233$}&\small{$-0.0100$}&\small{$-0.0376$}&\small{$0.0164$}\\
$\beta_M$&\small{$1.936$}&\small{$-0.915$}&\small{$-1.591$}&\small{$0.5033$}\\
$c_5$&\small{$0.790$}&\small{$-0.307$}&\small{$-0.576$}&\small{$0.2038$}\\
$c_6$&\small{$0.0874$}&\small{$-0.0239$}&\small{$-0.0750$}&\small{$0.0302$}\\
$s$&\small{$1.70$}&\small{$-0.8512$}&\small{$-1.268$}&\small{$2.277$}\\
$\rho$&\small{$0.399$}&\small{$-0.2357$}&\small{$-0.2682$}&\small{$0.4950$}\\
$\alpha$&\small{$-0.1257$}&\small{$0.02618$}&\small{$0.0457$}&\small{$-0.1811$}\\
$c_7$&\small{$-0.698$}&\small{$0.3518$}&\small{$0.5207$}&\small{$-0.874$}\\
$c_8$&\small{$-0.386$}&\small{$0.247$}&\small{$0.3316$}&\small{$-0.0765$}\\
$d$&\small{$-0.568$}& \small{$0.270$}&\small{$0.3955$}&\small{$-0.7605$}\\ \Xhline{1.2pt}
\end{tabular}
\label{SMTD}
\end{table}
\begin{table}
\caption{Sensitivity of the metronomic approach to model parameters.}
\centering
\begin{tabular} {c|c c c c}\Xhline{1.2pt}
\diagbox[width=10em]{\thead {Parameters}}{\thead {Types of \\cells}}&\thead {$N$}&\thead {$T_I$}&\thead {$T_M$}&\thead {$I$}\\ \Xhline{1.2pt}
$b_1$&\small{$1.866$}&\small{$-0.106$}&\small{$-0.0134$}&\small{$0.024$}\\
$c_1$&\small{$-1.86$}& \small{$0.1151$}&\small{$0.0827$}&\small{$-0.030$}\\
$c_2$&\small{$-0.951$}&\small{$0.0687$}&\small{$0.029$}&\small{$-0.0136$}\\
$r$&\small{$-3.609$}&\small{$1.983$}&\small{$2.237$}&\small{$-0.053$}\\
$b_2$&\small{$-3.126$}&\small{$5.019$}&\small{$2.077$}&\small{$-0.900$}\\
$\alpha_I$&\small{$-2.195$}&\small{$1.617$}&\small{$0.457$}&\small{$-0.659$}\\
$\alpha_M$&\small{$-1.017$}&\small{$0.075$}&\small{$2.259$}&\small{$-0.2806$}\\
$\beta_I$&\small{$0.6146$}&\small{$-0.4016$}&\small{$-0.4363$}&\small{$0.033$}\\
$c_3$&\small{$0.9664$}&\small{$-0.5855$}&\small{$-0.6162$}&\small{$0.293$}\\
$c_4$&\small{$0.0314$}&\small{$-0.0189$}&\small{$-0.0037$}&\small{$0.00899$}\\
$\beta_M$&\small{$1.6518$}&\small{$-0.798$}&\small{$-1.633$}&\small{$0.487$}\\
$c_5$&\small{$0.65$}&\small{$-0.238$}&\small{$-0.774$}&\small{$0.197$}\\
$c_6$&\small{$0.0729$}&\small{$-0.0955$}&\small{$-0.1302$}&\small{$0.02514$}\\
$s$&\small{$1.583$}&\small{$-0.848$}&\small{$-1.418$}&\small{$2.285$}\\
$\rho$&\small{$0.319$}&\small{$-0.217$}&\small{$-0.264$}&\small{$0.508$}\\
$\alpha$&\small{$-0.128$}&\small{$0.0778$}&\small{$0.0934$}&\small{$-0.1874$}\\
$c_7$&\small{$-0.573$}&\small{$0.388$}&\small{$0.477$}&\small{$-0.838$}\\
$c_8$&\small{$-0.309$}&\small{$0.227$}&\small{$0.262$}&\small{$-0.075$}\\
$d$&\small{$-0.512$}& \small{$0.328$}&\small{$0.399$}&\small{$-0.81$}\\ \Xhline{1.2pt}
\end{tabular}
\label{SMetronomic}
\end{table}

\subsection{Analysing the sensitivity indices }
A comparison of Table \ref{SMTD} and Table \ref{SMetronomic} reveals similar trends in the sensitivity of parameters across both treatment approaches, with minor discrepancies. The most sensitive parameters for each cell category are largely consistent between MTD and metronomic protocols, with slight variations. For example, for $T_I$ cells, the most sensitive parameters under MTD are $b_2,~\alpha_I,r,\beta_M,$ and $s$, while under metronomic treatment, they are $b_2,r,\alpha_I,~s,$ and $\beta_M$. Least sensitive parameters also show similar patterns, with differences in order.

Notably, $c_4$ is the least sensitive parameter in both approaches across all cell types, as it represents the competition between healthy and tumor cells during interphase, where healthy cells cannot compete effectively. Parameter $s$, representing the steady influx of immune system cells, is a major contributor to immune cell populations, highlighting its significant role in the host’s response to tumors and treatment.
Finally, parameters $b_2$ (carrying capacity) and $r$ (growth rate) are consistently among the most sensitive in all cell types, underscoring their critical roles in tumor evolution and chemotherapy response. The model's output is thus heavily influenced by tumor type.

\section{Conclusion}
In this study, a novel mathematical model based on delay differential equations was proposed to describe and predict the dynamics of cells in oncology, specifically breast cancer. Unlike previous models, which neglected tumor cell proliferation, the logistic growth function was employed to model tumor cell growth. Additionally, the model incorporates the competition among tumor, immune, and healthy cells. Parameter values were primarily derived from the biomathematics literature, and a history function was introduced due to the delay aspect of the model.
Through in silico experiments, clinical observations were mathematically embodied, offering rational explanations for these phenomena. The simulations also highlighted the importance of enhancing immune surveillance via adjacent therapies. The results demonstrate that the proposed model better aligns with biological tumor size oscillations, in contrast to previous models that unrealistically predicted tumor eradication by chemotherapy. The model’s consistency with experimental data suggests its robustness in predicting tumor response to various treatment protocols.
Theoretically, the model confirms that a healthy immune system can eradicate tumor cells, while a weakened immune system allows for continued tumor proliferation, underscoring the need for treatments that bolster immune function. This work provides valuable insights for future clinical applications and preclinical studies.

\section*{Funding}
This research received no external funding.

\section*{Conflict of Interest}
The authors declare that there are no conflicts of interest regarding the publication of this paper.

\bibliography{BibPhDThesis}

@Article{ansarizadeh2017modelling,
  author    = {Ansarizadeh, F. and Singh, M. and Richards, D.},
  title     = {Modelling of tumor cells regression in response to chemotherapeutic treatment},
  journal   = {Applied Mathematical Modelling},
  year      = {2017},
  publisher = {Elsevier},
}

@Article{colleoni2002low,
  author    = {Colleoni, M. and Rocca, A. and Sandri, M. T. and Zorzino, L. and Masci, G. and Nole, F. and Peruzzotti, G. and Robertson, C. and Orlando, L. and Cinieri, S. and others},
  title     = {Low-dose oral methotrexate and cyclophosphamide in metastatic breast cancer: antitumor activity and correlation with vascular endothelial growth factor levels},
  journal   = {Annals of Oncology},
  year      = {2002},
  volume    = {13},
  number    = {1},
  pages     = {73--80},
  publisher = {Oxford University Press},
}

@Article{costanza2008epidemiology,
  author  = {Costanza, M. E. and Chen, W. Y.},
  title   = {Epidemiology and risk factors for breast cancer},
  journal = {Available via www. uptodateonline. com. Accessed},
  year    = {2008},
  volume  = {10},
}

@Article{de2001mathematical,
  author    = {De Pillis, L. G. and Radunskaya, A.},
  title     = {A mathematical tumor model with immune resistance and drug therapy: {A}n optimal control approach},
  journal   = {Computational and Mathematical Methods in Medicine},
  year      = {2001},
  volume    = {3},
  number    = {2},
  pages     = {79--100},
  publisher = {Taylor \& Francis},
}

@Article{de2003dynamics,
  author    = {De Pillis, L. G. and Radunskaya, A.},
  title     = {The dynamics of an optimally controlled tumor model: {A} case study},
  journal   = {Mathematical and Computer Modelling},
  year      = {2003},
  volume    = {37},
  number    = {11},
  pages     = {1221--1244},
  publisher = {Elsevier},
}

@Article{de2006paradoxical,
  author    = {De Visser, K. E. and Eichten, A. and Coussens, L. M.},
  title     = {Paradoxical roles of the immune system during cancer development},
  journal   = {Nature reviews cancer},
  year      = {2006},
  volume    = {6},
  number    = {1},
  pages     = {24},
  publisher = {Nature Publishing Group},
}

@Article{enderling2007mathematical,
  author    = {Enderling, H. and Chaplain, M. A. and Anderson, A. R. and Vaidya, J. S.},
  title     = {A mathematical model of breast cancer development, local treatment and recurrence},
  journal   = {Journal of theoretical biology},
  year      = {2007},
  volume    = {246},
  number    = {2},
  pages     = {245--259},
  publisher = {Elsevier},
}

@Article{hanahan2000less,
  author    = {Hanahan, D. and Bergers, G. and Bergsland, E.},
  title     = {Less is more, regularly: metronomic dosing of cytotoxic drugs can target tumor angiogenesis in mice},
  journal   = {The Journal of clinical investigation},
  year      = {2000},
  volume    = {105},
  number    = {8},
  pages     = {1045--1047},
  publisher = {Am Soc Clin Investig},
}

@Article{hanahan2000hallmarks,
  author    = {Hanahan, D. and Weinberg, R. A.},
  title     = {The hallmarks of cancer},
  journal   = {cell},
  year      = {2000},
  volume    = {100},
  number    = {1},
  pages     = {57--70},
  publisher = {Elsevier},
}

@Article{kuznetsov1994nonlinear,
  author    = {Kuznetsov, V. A. and Makalkin, I. A. and Taylor, M. A. and Perelson, A. S.},
  title     = {Nonlinear dynamics of immunogenic tumors: parameter estimation and global bifurcation analysis},
  journal   = {Bulletin of mathematical biology},
  year      = {1994},
  volume    = {56},
  number    = {2},
  pages     = {295--321},
  publisher = {Elsevier},
}

@Book{niculescu2012advances,
  title     = {Advances in time-delay systems},
  publisher = {Springer Science \& Business Media},
  year      = {2012},
  author    = {Niculescu, S. I. and Gu, K.},
  volume    = {38},
}

@Article{richie2003breast,
  author  = {Richie, R. C. and Swanson, J. O.},
  title   = {Breast cancer: a review of the literature},
  journal = {JOURNAL OF INSURANCE MEDICINE-NEW YORK THEN DENVER--},
  year    = {2003},
  volume  = {35},
  number  = {2},
  pages   = {85--101},
}

@Article{42,
  author    = {J. Rieker},
  title     = {conversations, physician with pomona vally hospital, pomona, CA},
  year      = {1999},
  volume    = {28},
  number    = {12},
  pages     = {2543--2547},
  publisher = {Oxford Univ Press},
}

@Article{shochat1999using,
  author    = {Shochat, E. and HART, D. and AGUR, Z.},
  title     = {Using computer simulations for evaluating the efficacy of breast cancer chemotherapy protocols},
  journal   = {Mathematical Models and Methods in Applied Sciences},
  year      = {1999},
  volume    = {9},
  number    = {04},
  pages     = {599--615},
  publisher = {World Scientific},
}

@Article{tubiana1976comparison,
  author  = {Tubiana, M. and Malaise, E.},
  title   = {Comparison of cell proliferation kinetics in human and experimental tumors: response to irradiation.},
  journal = {Cancer treatment reports},
  year    = {1976},
  volume  = {60},
  number  = {12},
  pages   = {1887--1895},
}

@Article{villasana2004heuristic,
  author    = {Villasana, M. and Ochoa, G.},
  title     = {Heuristic design of cancer chemotherapies},
  journal   = {IEEE Transactions on Evolutionary Computation},
  year      = {2004},
  volume    = {8},
  number    = {6},
  pages     = {513--521},
  publisher = {IEEE},
}

@Article{wu2015time,
  author    = {Wu, L. and Lam, H. K. and Zhao, Y. and Shu, Z.},
  title     = {Time-delay systems and their applications in engineering 2014},
  journal   = {Mathematical Problems in Engineering},
  year      = {2015},
  volume    = {2015},
  publisher = {Hindawi},
}

@Article{7,
  title     = {American Brain Tumor Association, Providing and Pursuing Answers, www.abata.org.},
  volume    = {28},
  number    = {12},
  pages     = {2543--2547},
  publisher = {Oxford Univ Press},
}

@Article{9,
  title     = {Teratment for barin and spinal cord tumors},
  journal   = {Cancer Council Victoria, www.cancervic.org.au},
  year      = {2011},
  volume    = {28},
  number    = {12},
  pages     = {2543--2547},
  publisher = {Oxford Univ Press},
}

@Book{athawale2015chemotherapy,
  title     = {Chemotherapy appointment scheduling and operations planning},
  publisher = {The University of Akron},
  year      = {2015},
  author    = {Athawale, S.},
}

@Article{merlo2006cancer,
  author    = {Merlo, L. M. and Pepper, J. W. and Reid, B. J. and Maley, C. C.},
  title     = {Cancer as an evolutionary and ecological process},
  journal   = {Nature Reviews Cancer},
  year      = {2006},
  volume    = {6},
  number    = {12},
  pages     = {924},
  publisher = {Nature Publishing Group},
}

@Article{roe2011mathematical,
  author    = {Roe-Dale, R. and Isaacson, D. and Kupferschmid, M.},
  title     = {A mathematical model of breast cancer treatment with {CMF} and doxorubicin},
  journal   = {Bulletin of mathematical biology},
  year      = {2011},
  volume    = {73},
  number    = {3},
  pages     = {585--608},
  publisher = {Springer},
}

@InProceedings{wu2010sensitivity,
  author       = {Wu, W. H. and Wang, F. S. and Chang, M. S.},
  title        = {Sensitivity analysis of dynamic biological systems with time-delays},
  booktitle    = {BMC bioinformatics},
  year         = {2010},
  volume       = {11},
  pages        = {S12},
  organization = {BioMed Central},
}

@Article{zitvogel2008immunological,
  author    = {Zitvogel, L. and Apetoh, L. and Ghiringhelli, F. and Kroemer, G.},
  title     = {Immunological aspects of cancer chemotherapy},
  journal   = {Nature reviews immunology},
  year      = {2008},
  volume    = {8},
  number    = {1},
  pages     = {59},
  publisher = {Nature Publishing Group},
}

@Article{siegel2017cancer,
  author    = {Siegel, R. L. and Miller, K. D. and Jemal, A.},
  title     = {Cancer statistics, 2017},
  journal   = {CA: a cancer journal for clinicians},
  year      = {2017},
  volume    = {67},
  number    = {1},
  pages     = {7--30},
  publisher = {Wiley Online Library},
}

@Article{mcpherson2000abc,
  author    = {McPherson, K. and Steel, C. and Dixon, J. M.},
  title     = {{ABC} of breast diseases: breast cancer—epidemiology, risk factors, and genetics},
  journal   = {BMJ: British Medical Journal},
  year      = {2000},
  volume    = {321},
  number    = {7261},
  pages     = {624},
  publisher = {BMJ Publishing Group},
}

@PhdThesis{newbury2007numerical,
  author = {Newbury, G.},
  title  = {A numerical study of a delay differential equation model for breast cancer},
  school = {Virginia Tech},
  year   = {2007},
}

@article{sha2024global,
  title={Global burden of breast cancer and attributable risk factors in 204 countries and territories, from 1990 to 2021: results from the Global Burden of Disease Study 2021},
  author={Sha, Rui and Kong, Xiang-meng and Li, Xin-yu and Wang, Ya-bing},
  journal={Biomarker Research},
  volume={12},
  number={1},
  pages={87},
  year={2024},
  publisher={Springer}
}

@article{bhardwaj2024male,
  title={Male Breast Cancer: a Review on Diagnosis, Treatment, and Survivorship},
  author={Bhardwaj, Prarthna V and Gupta, Shilpi and Elyash, Alexa and Teplinsky, Eleonora},
  journal={Current Oncology Reports},
  volume={26},
  number={1},
  pages={34--45},
  year={2024},
  publisher={Springer}
}

@article{cazzaniga2021metronomic,
  title={Metronomic chemotherapy},
  author={Cazzaniga, Marina Elena and Cordani, Nicoletta and Capici, Serena and Cogliati, Viola and Riva, Francesca and Cerrito, Maria Grazia},
  journal={Cancers},
  volume={13},
  number={9},
  pages={2236},
  year={2021},
  publisher={MDPI}
}

@article{aprile2021hypertermic,
  title={Hypertermic Intrathoracic Chemotherapy (HITHOC) for thymoma: a narrative review on indications and results},
  author={Aprile, Vittorio and Bacchin, Diana and Korasidis, Stylianos and Ricciardi, Roberta and Petrini, Iacopo and Ambrogi, Marcello Carlo and Lucchi, Marco},
  journal={Annals of Translational Medicine},
  volume={9},
  number={11},
  year={2021},
  publisher={AME Publications}
}

@article{bandini2024metronomic,
  title={Metronomic Chemotherapy in Elderly Patients},
  author={Bandini, Arianna and Calabr{\`o}, Pasquale Fabio and Banchi, Marta and Orlandi, Paola and Bocci, Guido},
  journal={Current Oncology Reports},
  volume={26},
  number={4},
  pages={359--376},
  year={2024},
  publisher={Springer}
}

@article{basar2024optimizing,
  title={Optimizing cancer therapy: a review of the multifaceted effects of metronomic chemotherapy},
  author={Basar, Oyku Yagmur and Mohammed, Sawsan and Qoronfleh, M Walid and Acar, Ahmet},
  journal={Frontiers in Cell and Developmental Biology},
  volume={12},
  pages={1369597},
  year={2024},
  publisher={Frontiers Media SA}
}

@article{alonso2024decoding,
  title={Decoding the Nucleolar Role in Meiotic Recombination and Cell Cycle Control: Insights into Cdc14 Function},
  author={Alonso-Ramos, Paula and Carballo, Jes{\'u}s A},
  journal={International Journal of Molecular Sciences},
  volume={25},
  number={23},
  pages={12861},
  year={2024},
  publisher={MDPI}
}

@article{xu2021technological,
  title={Technological advances in cancer immunity: from immunogenomics to single-cell analysis and artificial intelligence},
  author={Xu, Ying and Su, Guan-Hua and Ma, Ding and Xiao, Yi and Shao, Zhi-Ming and Jiang, Yi-Zhou},
  journal={Signal Transduction and Targeted Therapy},
  volume={6},
  number={1},
  pages={312},
  year={2021},
  publisher={Nature Publishing Group UK London}
}

@article{liao2025inequality,
  title={Inequality in breast cancer: Global statistics from 2022 to 2050},
  author={Liao, Ling},
  journal={The Breast},
  volume={79},
  pages={103851},
  year={2025},
  publisher={Elsevier}
}

@article{matthews2022cell,
  title={Cell cycle control in cancer},
  author={Matthews, Helen K and Bertoli, Cosetta and de Bruin, Robertus AM},
  journal={Nature reviews Molecular cell biology},
  volume={23},
  number={1},
  pages={74--88},
  year={2022},
  publisher={Nature Publishing Group UK London}
}

@inproceedings{paul2022tumor,
  title={Tumor glycolysis, an essential sweet tooth of tumor cells},
  author={Paul, Sumana and Ghosh, Saikat and Kumar, Sushil},
  booktitle={Seminars in cancer biology},
  volume={86},
  pages={1216--1230},
  year={2022},
  organization={Elsevier}
}

@article{awang2022tumor,
  title={Tumour-natural killer and cD8+ t cells interaction model with delay},
  author={Awang, Nor Aziran and Maan, Normah and Sulain, Mohd Dasuki},
  journal={Mathematics},
  volume={10},
  number={13},
  pages={2193},
  year={2022},
  publisher={MDPI}
}
\bibliographystyle{ieeetr}

\end{document}